\documentclass[twocolumn,preprintnumbers,amsmath,amssymb,aps,pre,reprint,longbibliography]{revtex4-2}
\usepackage{graphicx}
\usepackage{xcolor}
\begin{document}

\title{
Phase Time Crystals and Pairing in Binary Active Chiral Systems 
}
\author{
C. Reichhardt and C. J. O. Reichhardt 
} 
\affiliation{Theoretical Division and Center for Nonlinear Studies,
Los Alamos National Laboratory, Los Alamos, New Mexico 87545, USA
}

\date{\today}

\begin{abstract}
We introduce a class of dynamic systems we call phase time crystals consisting of a binary assembly of particles with intermediate or long-range repulsive interactions that are subjected to a circular drive of uniform chirality in which each particle species is out of phase from the other by $180^\circ$. As a function of the particle density and orbit radius, this system can organize into a rich variety of dynamical crystalline states, including one in which the out of phase particles form bound pairs that assemble into a triangular lattice. We also find stripe phases, overlapping packed crystals, disordered or phase glass states with no diffusion, mixed fluids, and different types of phase-separated states. We show that these states are robust against the addition of thermal fluctuations, and that the paired crystal can melt into a paired fluid. If the drive on each particle species is of opposite chirality, the system forms stripes and packed lattices, but no paired crystal is present. We demonstrate that by modifying the nature of the chiral driving, it is possible to realize numerous kinds of active molecular lattices, including dynamic square spin ice geometries and higher-order complex structures.  
\end{abstract}

\maketitle

\section{Introduction}

In equilibrium, two-dimensional systems of particles with intermediate or long-range repulsive interactions of Bessel function, Yukawa, or Coulomb
type will form a triangular lattice. This process has been studied for
charged colloids \cite{Murray87,Pertsinidis01a,Libal07}, magnetically interacting colloids \cite{Zahn99}, vortices in type-II superconductors \cite{Guillamon09}, magnetic skyrmions \cite{Yu10}, Wigner crystals \cite{Tsui24}, and dusty plasmas \cite{Thomas94}.
An open question is what would happen to such structures if a time-dependent component were added to the interaction.
The system may 
still form a triangular lattice or fluid,
but it is also possible that dynamic patterns
could emerge
in which the time dependence creates
effective particle-particle interactions that are both attractive and repulsive.

One of the simplest examples of additional time-dependent interactions
is the introduction of
circular or chiral particle motion.
Recently there has been growing interest in
chiral active matter
systems where
the particles undergo circular motion or
are themselves spinning
\cite{Lemelle10,Kummel13,Nguyen14,Lowen16,vanZuiden16,Banerjee17,Han17,Soni19,Lei19,Reichhardt19,Liebchen22,Tan22,Baconnier19,Fruchart23,Gelvan25,Caprini26}.
Systems of active spinning particles
can form large-scale rotating solids \cite{Tan22}
with edge currents \cite{vanZuiden16,Soni19}, and can also exhibit
odd viscosity and odd elasticity effects when
the chirality produces transverse forces
\cite{Banerjee17,Soni19,Han21,Reichhardt22b,Fruchart23,Huang25}.
Mixtures of particles that have opposite active chirality can phase separate
or
form mixed fluids, with
edge currents appearing along the boundaries between the two species
\cite{Nguyen14,Han17,Reichhardt19,Massana-Cid21,Bililign22,Guo25}.
In other 
active chiral systems, the particles undergo circular motion
\cite{Han17,Reichhardt19,Zhang22,Jeong25}. Assemblies of
chiral robots can organize into rotating patterns
or other patterns \cite{Wang25,Kiechl26},
and it has even been shown how to use active rotation to achieve
self-assembly \cite{Aubret18}.
Most studies of chiral systems
have been performed with disks or particles
that have short-range or contact repulsion,
so that a crystal can form only in the dense limit,
but there has been some work on
active particles with Lennard-Jones
type interactions \cite{Caporusso24} or attractive interactions \cite{Li24}.
Less is known about how chiral and non-chiral active systems
would behave if the particles have intermediate or long-range repulsion;
however,
there are growing efforts to create
active matter systems with
this type of interactions
\cite{Massana-Cid21,Massana-Cid24,Sungar25,Vyas26},
including recent proposals for
magnetic skyrmion systems with activity \cite{Silva25}.
In general, if the repulsive interactions are
sufficiently strong, it could be expected that the system could simply
form a triangular solid by overcoming any active forces.
If the activity is stronger, however,
the system could remain fluid, rotate as a solid,
or produce new types of ordered structures as a result of
the competition between repulsion and chirality.

\begin{figure}
\includegraphics[width=\columnwidth]{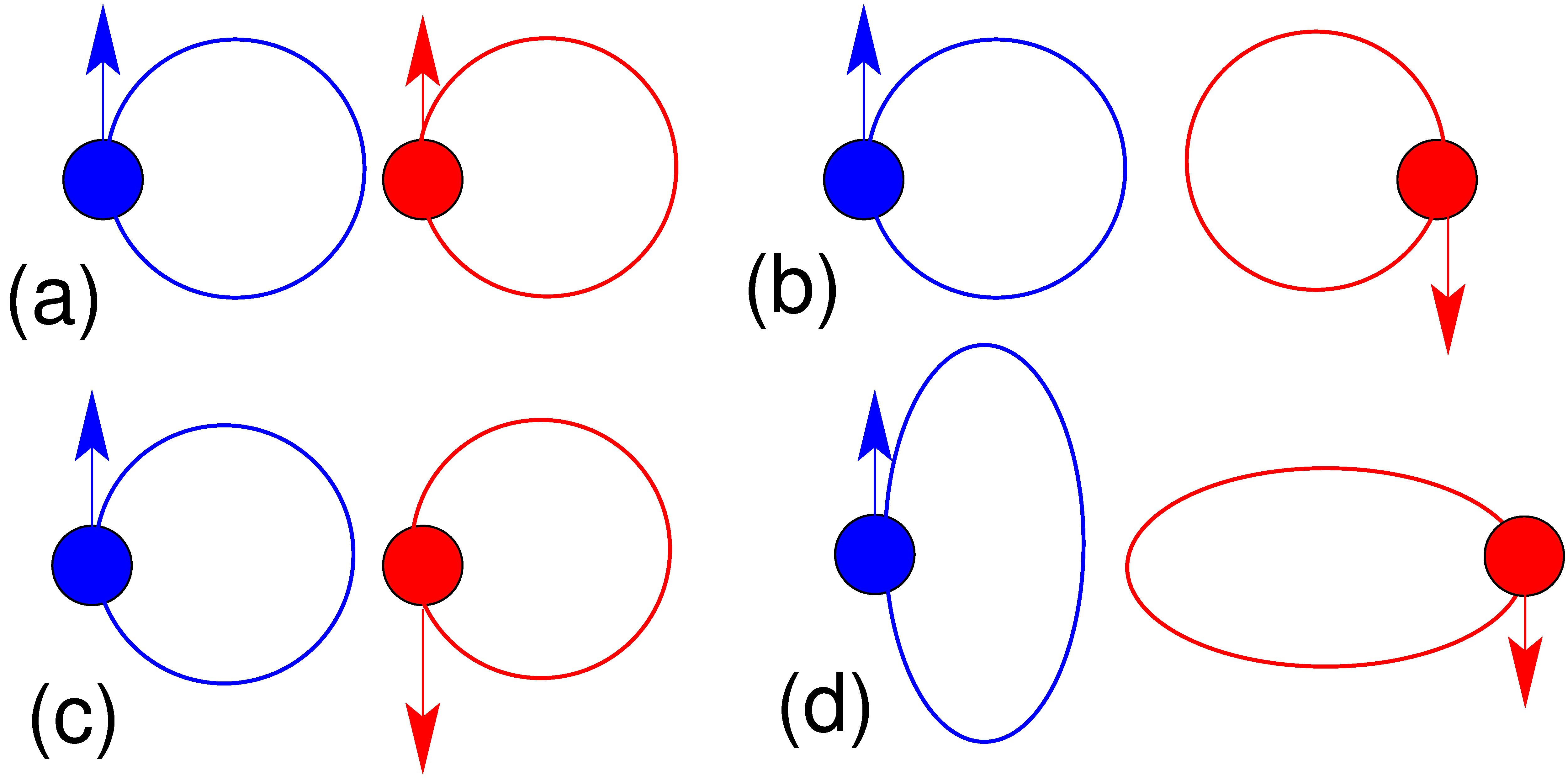}
\caption {Schematics illustrating active chiral driving
for a binary system of particles with intermediate or long-range repulsion.
(a) Two particles that both rotate in the same direction while in phase.
(b) The two particles rotate in the same direction
but are out of phase by $180^\circ$.
(c) Particles that rotate with opposite chirality.
(d) The particles rotate with the same chirality,
are out of phase by $180^\circ$,
and have different drive amplitudes in the $x$ and $y$ directions.
}
\label{fig:1}
\end{figure}

In this work we propose that phase time crystal states
can appear in
an active chiral system
consisting of a binary assembly of particles with
intermediate or long-range repulsion.
When the chiral driving is out of phase for the two particle species,
it can induce an effective attraction
between the particles.
In addition to triangular lattices, this system
forms paired crystals, stripe lattices,
packed overlapping crystals,
and glassy states, along with phase-separated and mixed fluids.
We demonstrate these states for particles with Bessel function,
Yukawa, and Coulomb repulsive interactions.
In each case the particles form a simple triangular lattice in equilibrium,
but
we add a chiral driving force
such that an individual particle
travels in a circle with radius $R_a$ at frequency $\omega$.
For a fixed driving amplitude, the orbit radius
increases as $\omega$ decreases.
When the particles are all driven with the same chirality and phase,
as shown in Fig.~\ref{fig:1}(a),
the system forms a triangular solid that rotates rigidly
so that in a rotating reference frame,
the particles are in a stationary triangular lattice.
When we consider a binary assembly of particles 
rotating with the same chirality but
out of phase by $180^\circ$ with each other,
as shown in Fig.~\ref{fig:1}(b),
the particles organize into multiple distinct
phases depending on the value of $\omega$ and the particle density.
For high $\omega$, when $R_a$ is much smaller
than the average lattice spacing $a$ of the equilibrium system,
the particles form a triangular lattice with non-overlapping orbits.
As the drive frequency decreases, square lattices or disordered states
can form instead.
At lower frequencies when the active radius $R_a$ is
close to the equilibrium lattice spacing $a$,
the system forms bound pairs of particles in which
two particles that are out of phase rotate around their
mutual center of mass, and the bound pairs themselves form a lattice.
For other frequencies, the system can form
phase-separated states with or without edge currents.
When the active radius is larger than the equilibrium lattice spacing,
the system can in some cases form overlapping
tightly packed crystals as well as stripe crystal structures,
and in other cases passes through a transient fluctuating state
before settling into a dynamically disordered frozen state with
no long-time diffusion.
These states are robust against
the inclusion of thermal fluctuations.
For the paired crystal lattice, as the temperature increases
the system first melts into a
paired fluid where the particles remain bound
into pairs of opposite phase, but the pairs can gradually diffuse.
At higher temperatures, the pairs break apart and a phase-separated state
emerges, while at the highest temperatures
we find a mixed fluid state.
At zero temperature, when the system is initialized
in a disordered state, it can organize
into a frozen mixture of pairs and stripes.
For a binary system with opposite chirality,
shown in Fig.~\ref{fig:1}(c),
we find stripe lattices, overlapping crystals,
and glass states,
but the paired crystal state is absent.
In the case of elliptical driving,
illustrated in Fig.~\ref{fig:1}(d),
the system can still form
cross-like paired crystals
as well as more complex lattices,
including states similar to those found for
square spin ice \cite{Nisoli13,Ambriz19} and superlattice orderings.

The phase time crystal system we propose
can be viewed as an extension of the time crystal concept.
Time crystals were originally proposed by Wilczek as systems in
which the ground state is periodic in both space and time
\cite{Wilczek12,Shapere12}, but it was soon shown that it is not possible
for time crystals to form
in equilibrium \cite{Bruno13}.
When nonequilibrium effects are added,
however, various types of time crystals can occur,
including versions that are purely classical \cite{Yao20,Nicolaou21,Zhao25}
or produced by non-reciprocal interactions \cite{Morrell26}.
More recently, time crystals have been proposed
to occur for active matter systems with chiral interactions \cite{Liu26}.
It has also been suggested that
particles with Yukawa interactions on a dynamic substrate \cite{Libal20} and
magnetic colloids on patterned substrates \cite{Ernst22}
could act as time-crystal-like systems.
The system we describe in the present work
could be realized for charged or magnetic colloidal assemblies
where there is an additional chiral activity term,
or using mixtures of charged particles
where half of the particles also have magnetic interactions.
It could also be realized using
binary magnetic particle mixtures, such as skyrmion-skyrmionium mixtures,
that have intermediate-range repulsion but
respond differently to an oscillating field, or in
charged or magnetic systems with local driving.
Our results suggest new methods for self assembly
and for creating an effective attraction
between particles whose equilibrium interactions are
purely pairwise repulsive.

\section{Simulation}

We model a two-dimensional system of size $L \times L$ with
$L=36$ to $L=64$ that has
periodic boundary conditions in the $x$- and $y$-directions.
The system contains $N$ particles at a density of $\rho=N/L^2$,
and is evenly divided between two particle species,
with $N/2$ particles of species A and $N/2$ particles of species B.
The particles have repulsive interactions, independent of
species type, and
are initially placed in a triangular lattice.
We first consider particles with intermediate length repulsion represented
by a modified Bessel function, $V(R_{ij}) = F_0K_{0}(R_{ij})$,
where $F_0$ is a constant and ${\bf R}_{i(j)}$ is the position of
particle $i(j)$.
The Bessel function decays exponentially at large distances 
and describes interactions between vortices in type-II superconductors,
skyrmions in chiral magnets, and is also a good approximation
for a variety of screened-charge systems \cite{Reichhardt17}.
We consider two other types of interaction potentials and find that they
give similar results. The first is a
repulsive Yukawa potential, $V(R_{ij}) = C e^{-R_{ij}}/R_{ij}$, where $C$ is a
constant prefactor.
The second is
a long-range Coulomb potential, $V(R_{ij}) = {Q}/{R_{ij}}$,
and in this case we employ the Lekner summation technique for
computational efficiency \cite{Lekner91,GronbechJensen97a},
as used previously for driven
charge motion on random substrates \cite{Reichhardt22}.
If thermal fluctuations are sufficiently small,
particles with any of these interaction
potentials will form a triangular lattice.
We introduce active chiral driving terms ${\bf F}^A$ and ${\bf F}^B$ that
cause individual particles of species A or B, respectively, to move in a
circle of radius $R_a$ at a frequency $\omega$. 

The overdamped equation of motion for particle $i$ is 
\begin{equation} \eta \frac{d {\bf R}_i}{dt} = -\sum_{j\neq i}^{N}\nabla V(R_{ij}) + {\bf F}^{A}_{i}\delta(\sigma_i-1) + {\bf F}^B_{i}\delta(\sigma_i)
\end{equation} 
where the damping coefficient is set to $\eta = 1.0$
and
particles of species A are assigned $\sigma_i=1$ while particles of
species B are assigned $\sigma_i=0$.
In the initial triangular lattice,
the particles are placed such that
every other lattice site
is occupied by a species A or a species B particle to give maximal mixing,
where the A and B particle species can
be subjected to different chiral driving protocols
as shown in Fig.~\ref{fig:1}.
For the main part of this work, we consider the protocol in
Fig.~\ref{fig:1}(b) where the particles have the same chirality and
same driving amplitude but are out of phase by $180^{\circ}$. 
In this case,
species A is subjected to the rotating drive
${\bf F}^{A}(t)=A\sin(\omega t)\hat{\bf y}+A\cos(\omega t)\hat{\bf x}$, while
species B has a rotating drive of
${\bf F}^{B}(t)= -A\sin(\omega t)\hat{\bf y} - A\cos(\omega t)\hat{\bf x}$.
In Fig.~\ref{fig:1}(a) where ${\bf F}^A={\bf F}^B$,
the system effectively contains only one particle species and all of
the particles rotate as a coherent solid.
When the particles are driven with opposite chirality,
as in Fig.~\ref{fig:1}(c),
species A has the 
same drive as above
but the species B drive is changed to
${\bf F}^{B}(t)= A\cos(\omega t)\hat{\bf y} + A\sin(\omega t)\hat{\bf x}$.
Unless otherwise stated, the ac drive amplitude is set to $A=1.0$.
We also consider the case
shown in Fig.~\ref{fig:1}(d) where the $x$ and $y$ driving amplitudes have
different values
to create elliptical orbits.

Throughout this work, we use a simulation time step of size
$\delta t = 0.005$,
and we report time in units of the period $\tau=2\pi/\omega$ of the ac drive.
We measure the fraction of sixfold-coordinated particles without regard to
species identity, $P_{6}= (1/N)\sum_{i=1}^{N}\delta(z_{i}-6)$, where $z_{i}$ is
the coordination number of particle $i$ obtained from a Voronoi tessellation.
For a triangular lattice, $P_6 = 1.0$. 
We also measure the mean square displacement versus
time per particle, $d(t) = (1/N)\sum^{N}_{i=1}|R_i(t) -R_i(t_{0})|^2$,
where $R_i(t)$ is the position of particle $i$ at time $t$
and the measurement is made from a reference time $t=t_0$.
In a fluid, $d(t)$ grows linearly with time,
while for a crystal or frozen state, $d(t)$ saturates to a constant value.

\section{Chiral System Out of Phase}

\begin{figure}
\includegraphics[width=\columnwidth]{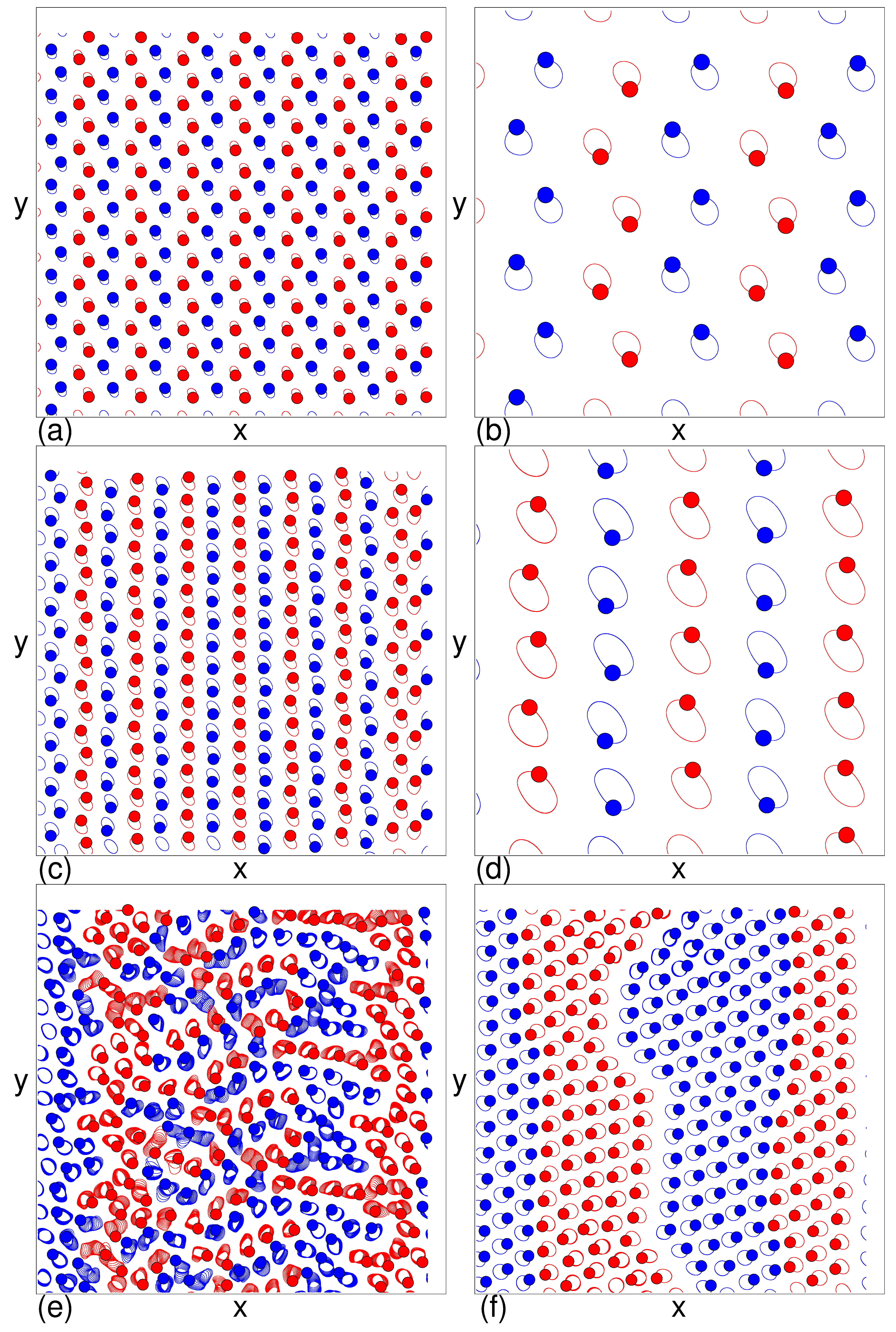}
\caption{Particle positions (circles) and trajectories (lines)
for Bessel function particles at density $\rho=0.208$
where species A is blue and species B is red.
The particles are subjected to circular driving
of the same chirality but are out of phase by $180^\circ$
as shown in Fig.~\ref{fig:1}(b); the drive amplitude is
$A=1.0$.
Except where stated,
the trajectories are obtained during an interval of 20 to 50 ac drive
cycles after the system has been allowed to evolve for
$10^4$ simulation time steps.
(a) A distorted triangular
lattice (phase I) at $\omega = 0.012$.
(b) A blow-up of panel (a)
showing more clearly that the orbits are slightly elliptical.
(c) At $\omega = 0.008$ the system forms a stripe lattice
(phase II) with
more elliptical orbits.
(d) A blow-up of panel (c).
(e) At early times for
$\omega = 0.0075$, the system breaks up into a transient fluid.
(f) The same as panel (e) after $5\times 10^4$
ac driving cycles have elapsed and
the system has settled into a phase-separated solid (phase III).
}
\label{fig:2}
\end{figure}

We first consider the situation from
Fig.~\ref{fig:1}(b) where particles with Bessel function interactions have
the same direction of chirality,
but the two species are out of phase by $180^\circ$.
We begin with the high frequency limit
where the orbit radius is much smaller than the average particle spacing.
In Fig.~\ref{fig:2}(a,b) we illustrate the particle configurations and
trajectories
for a system with $\rho = 0.208$ at $\omega = 0.012$.
For this density, the non-driven system
forms a triangular lattice with a spacing of $a = 2.19$, while
the particle orbits at $\omega=0.012$
have an average radius of $R_a=0.4$, so there is no overlap of
adjacent orbits.
Before taking the images in Fig.~\ref{fig:2}(a,b),
we wait $1\times 10^4$ simulation time steps, which
is sufficient time for any initial transients to
subside and for the system to settle into a steady cyclic state.
We illustrate the trajectories
during 20 to 50 ac drive cycles.
Here the steady state consists
of a distorted triangular lattice of monomers of rotating particles
(phase I).
In Fig.~\ref{fig:2}(b) we show a
zoomed in view of a smaller subsection of the system
that more clearly indicates that the two species are out of phase with
each other in their orbits, which
are not completely circular but are somewhat elliptical.
The slight noncircularity of the orbits is a result of
the particle-particle interactions:
since the particles are out of phase with each other,
they can move slightly further apart from each other
and minimize their interaction energy by distorting the
orbit shapes.
At higher values of $\omega$ (not shown),
we obtain the same structure as in Fig.~\ref{fig:2}(a,b),
but the orbits become smaller and more circular,
and the triangular lattice
formed by the particles becomes less distorted.
The particles repeat the exact same trajectory during each
drive cycle, so there is no diffusion.
The type of ordering we observe can have some dependence
on the initial particle placement, as discussed in a later section.
When we lower the driving frequency, the orbit size increases
and the system settles into different monomer patterns,
as shown in Fig.~\ref{fig:2}(c,d) at $\omega = 0.008$,
where the orbits are more noticeably elliptical
and the overall ordering can be described as a stripe lattice (phase II). All of
the ellipses for the orbits of both particle species are tilted in the
same direction.
At even lower frequencies, the elliptical orbits become
increasingly elongated and the system breaks up into a
transient disordered fluid phase,
illustrated in Fig.~\ref{fig:2}(e)
for $\omega = 0.0075$.
Over time the particles organize
into a phase separated solid (phase III),
shown in Fig.~\ref{fig:2}(f) for the $\omega=0.0075$ system
after several thousand ac driving cycles
have elapsed.
In the phase separated state, each particle species
forms a triangular lattice, the orbits become much more
circular, and there is no long-time diffusion.

\begin{figure}
\includegraphics[width=\columnwidth]{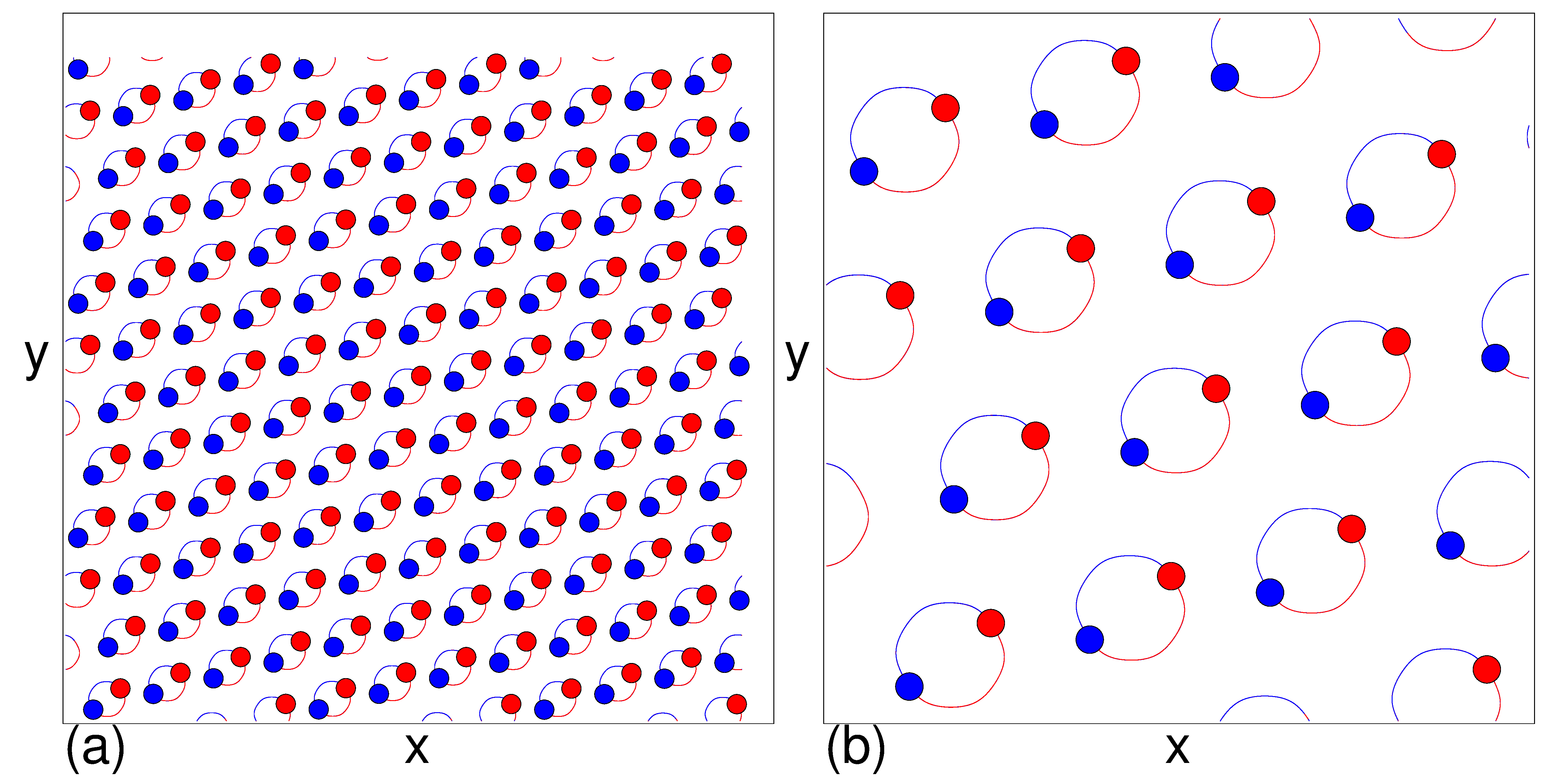}
\caption {(a) Particle positions (circles) and trajectories (lines)
where species A is blue and species B is red
for the
out of phase Bessel function particles at $\rho=0.208$ from Fig.~\ref{fig:2}
in the paired crystal state (phase IV) at $\omega = 0.0055$,
where the particles share orbits as pairs and these pairs then form
a higher-order crystal. (b) A blow-up of panel (a).
}
\label{fig:3}
\end{figure}

Over the range $0.0045 < \omega < 0.007$ at the same density of $\rho=0.208$
as in Fig.~\ref{fig:2},
we find an interesting paired crystal state (phase IV).
Figure~\ref{fig:3}(a) shows an 
example of the paired crystal at $\omega = 0.0055$.
Pairs of particles that are out of phase share a single orbit,
and these pairs then form a larger scale triangular lattice. The
zoomed in view of a portion of the sample in
Fig.~\ref{fig:3}(b) illustrates more clearly that
the paired particles are in the same orbit
but are out of phase with each other, and are traveling
around a center of mass point.
The range of $\omega$ over which
the paired crystal can form
depends on $\rho$ and $A$.
At early times, the system
can form a square lattice of pairs that then undergo a series of
transient rearrangements before settling into a steady state
triangular lattice.
The paired crystal state can emerge
when the orbit radius $R_a$ becomes large enough that two adjacent
orbits would overlap;
however, by sharing the same orbit, the two particle species
can form a lattice containing orbits
that are smaller than the lattice constant of the paired state.
Since the lattice constant goes as
$a \propto \rho^{-1/2}$,
and the paired particle density is half that
of the unpaired particle,
the lattice of the paired state is larger by
$\sqrt{2}$ than that of the unpaired state.
For larger $\omega$ in the paired state,
the orbits exhibit
small distortions away from being circular.

\begin{figure}
\includegraphics[width=\columnwidth]{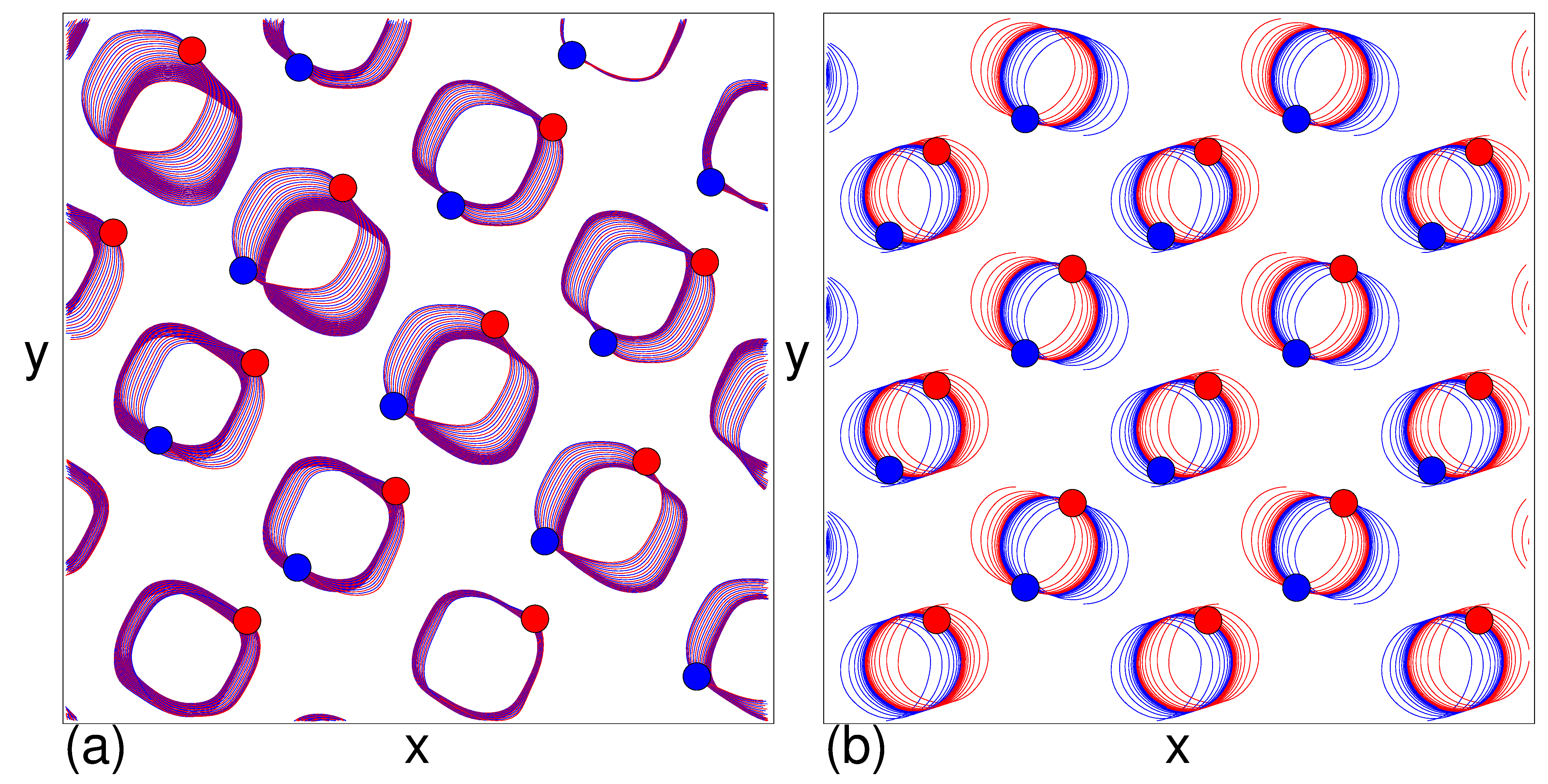}
\caption {Particle positions (circles) and trajectories (lines)
where species A is blue and species B is red
for the out of phase
Bessel function particles at $\rho=0.208$ from Fig.~\ref{fig:2} in
the paired crystal state (phase IV).
(a) At early times for $\omega=0.005$ the orbit shape has a square
character. After some rearrangements occur,
the system settles into a triangular lattice with nearly circular orbits.
(b) Early time evolution at $\omega=0.005$
for weaker particle-particle interactions of $F_{0} = 0.2$,
showing that there is an effective attraction between particles that
brings the particles together into paired orbits.
}
\label{fig:4}
\end{figure}

In Fig.~\ref{fig:4}(a) we show that at early times in
the paired crystal state at $\omega = 0.005$, the orbits have
a more square shape.
After a transient period,
the system rearranges and settles into a triangular paired lattice.
Figure~\ref{fig:4}(b) shows the early time evolution for a system
still at
$\omega = 0.005$ but with a reduced
particle-particle interaction strength of $F_{0} = 0.2$.
The out of phase
particles are initially in separate orbits but are effectively attracted
to each other in pairs, forming the paired crystal at later time.

\begin{figure}
\includegraphics[width=\columnwidth]{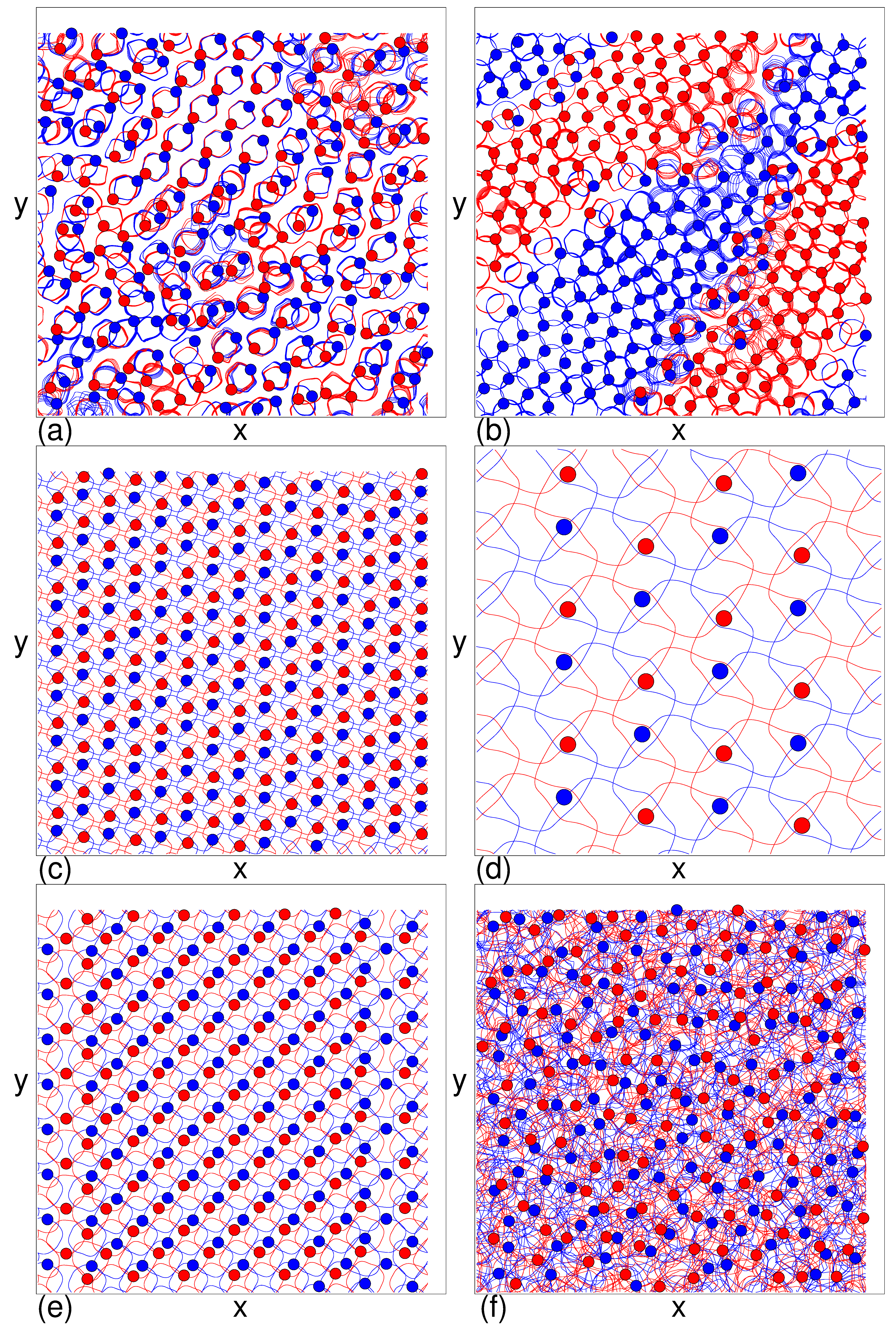}
\caption{Particle positions (circles) and trajectories (lines) where
species A is blue and species B is red for the out of phase
Bessel function particles
at $\rho=0.208$ from Fig.~\ref{fig:2}.
(a) The early time configuration and trajectories at $\omega = 0.004$
showing fluid motion.
(b) The steady state configuration for $\omega=0.004$
where the system forms a phase-separated fluid (phase V).
(c) An overlapping packed crystal (phase VI) at $\omega = 0.0025$.
(d) A blow-up of panel (c).
(e) The overlapping packed crystal at $\omega = 0.00325$,
where the shape of the orbit is different compared to panels (c, d).
(f) The mixed fluid state (phase VII) for $\omega = 0.002$.
}
\label{fig:5}
\end{figure}

For $0.00325 \leq \omega \leq 0.0045$, the
particle orbits become so large that the paired crystal is not able to form
and instead 
a phase-separated state emerges;
however, since the orbits are large,
the system has a fluid-like behavior with edge currents
flowing along the boundaries between the two phases.
In Fig.~\ref{fig:5}(a) we show the early time configurations
and trajectories at $\omega = 0.004$,
where there is some pairing of particles
but there is fluid-like motion throughout the system.
At later times in the steady state,
Fig.~\ref{fig:5}(b) shows that a phase-separated state
has emerged,
but the particle
orbits are larger than the average spacing between particles
and there is considerable motion throughout the system.
This phase-separated fluid (phase V) contains some local triangular ordering,
and the particles can gradually move to the phase separation boundaries where
a slight amount of mixing of particle species occurs.
For $0.00235 \leq \omega \leq 0.00325$, we find
an ordered overlapping packed crystal state (phase VI),
as shown in Fig.~\ref{fig:5}(c) for $\omega = 0.0025$.
The close-up view of this state in
Fig.~\ref{fig:5}(d) indicates that
even though the particle orbits are larger than
the average particle spacing,
the system can still form a dynamically ordered state.
The overlapping crystal state at a slightly higher frequency
of $\omega=0.003$ 
in Fig.~\ref{fig:5}(e) has orbits of altered shape compared to
the $\omega=0.0025$ system in Fig.~\ref{fig:5}(c,d).
For $\omega < 0.00235$, the system forms a mixed fluid (phase VII),
illustrated at $\omega=0.002$ in Fig.~\ref{fig:5}(f).
This mixed fluid state does not phase separate even at long times.
For much lower driving frequencies,
the system can form laned states, which will be studied in a different work.

\begin{figure}
\includegraphics[width=\columnwidth]{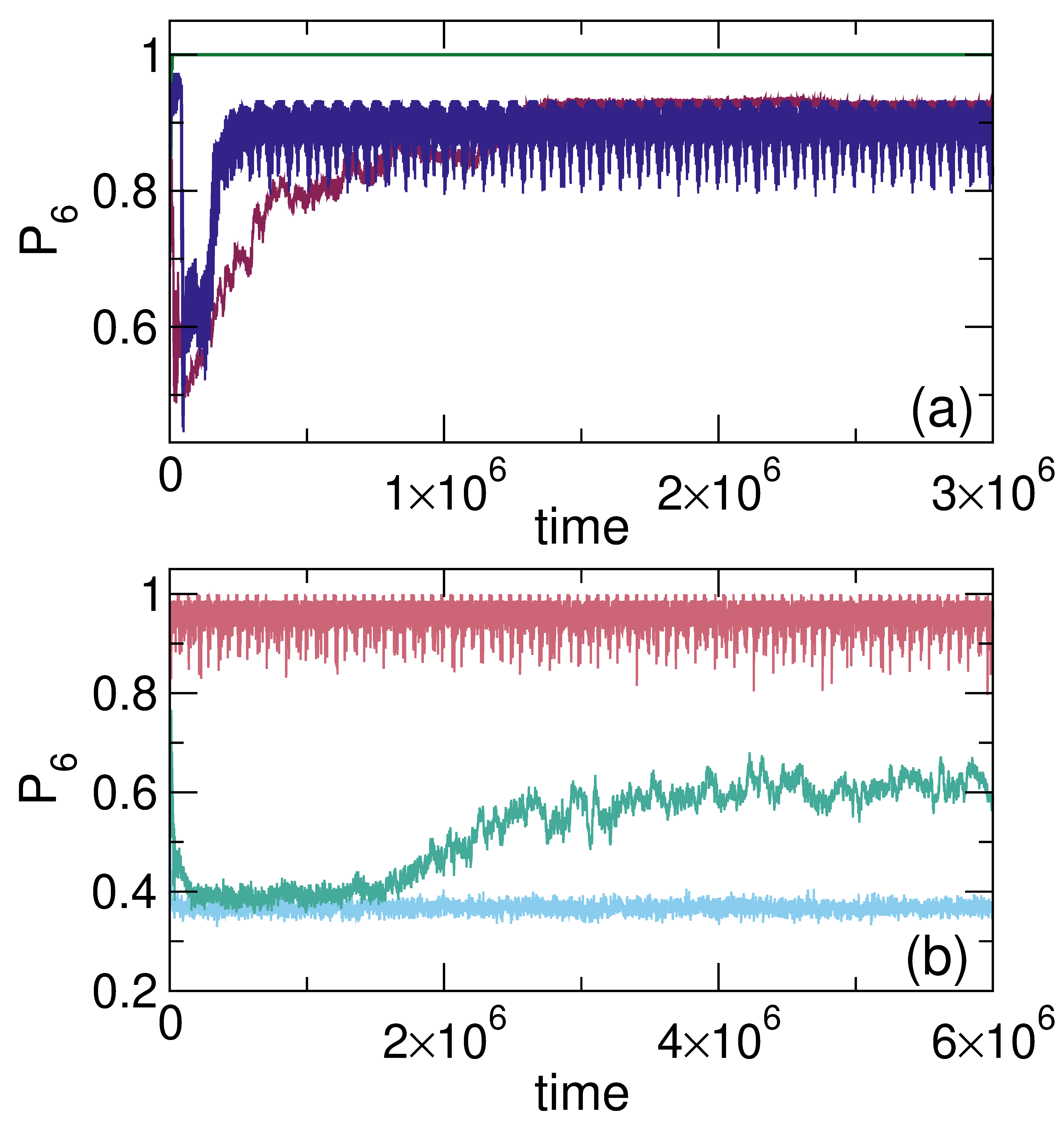}
\caption{$P_6$ versus time for the out of phase
Bessel function particles at $\rho=0.208$
from Fig.~\ref{fig:2}.
(a) Green: distorted triangular lattice (phase I) at $\omega = 0.012$.
Purple: $\omega = 0.0075$ where
the system forms a fluid that
gradually phase separates into a phase separated solid (phase III).
Dark blue: paired crystal (phase IV) at $\omega = 0.0055$.
(b) Teal: phase separated fluid (phase V) at $\omega = 0.004$.
Pink: overlapping packed crystal (phase VI) at $\omega = 0.0025$.
Light blue: mixed fluid (phase VII) at $\omega = 0.002$.
}
\label{fig:6}
\end{figure}

To get a better idea of the dynamics,
in Fig.~\ref{fig:6}(a) we plot $P_6$ versus time for the system
from Figs.~\ref{fig:2} to \ref{fig:5} at three values of $\omega$.
For $\omega = 0.012$, $P_6 \approx 1.0$ and the
system immediately forms a distorted triangular lattice (I)
as shown in Fig.~\ref{fig:2}(a,b).
At $\omega = 0.0075$, initially the particles disorder and have
$P_6 \approx 0.5$ in the transient fluid state shown in Fig.~\ref{fig:2}(e),
but $P_6$ increases over time and saturates to
$P_6\approx 0.9$ in the phase-separated state shown in Fig.~\ref{fig:2}(f).
The system does not completely order
in the phase-separated state due to the presence of
topological defects on the boundary separating the two phases.
In general, this state can be described as a phase-separated solid (III).
For $\omega = 0.0055$, after an initial period of transient motion
the system settles into a paired crystal state (IV) with $P_6 = 0.88$.
The Voronoi construction is based on the instantaneous locations of
the individual particles,
and in the paired crystal state,
$P_6$ is slightly depressed during a portion of each cycle, causing the
average value of $P_6$ to be noticeably lower than $1.0$ even though
a Voronoi construction performed on the center of mass of each pair
would show $P_6 \approx 1.0$.

In Fig.~\ref{fig:6}(b), we
plot $P_{6}$ versus time for the disordered
or phase separated fluid state (V) at $\omega = 0.004$.
At early times 
the system starts off in a mixed fluid state with
$P_6 \approx 0.4$, as shown in Fig.~\ref{fig:5}(a),
but then becomes better phase separated, as illustrated in
Fig.~\ref{fig:5}(b), causing $P_6$ to increase
to $P_6 \approx 0.6$.
This phase-separated state is not as well ordered
as the phase-separated solid
that appears for $\omega = 0.0075$,
and it is better described as a phase-separated fluid.
For $\omega = 0.0025$ in the overlapping packed crystal state (VI)
shown in Fig.~\ref{fig:5}(c,d),
$P_6$ is high again.
The fluctuations in $P_6$ are produced when the particles approach each other
more closely during portions of the ac cycle, disrupting the order locally.
In the mixed fluid (VII) from Fig.~\ref{fig:5}(f) at $\omega = 0.002$,
$P_6$ is close to $P_6=0.39$ and does not change with time.

\begin{figure}
\includegraphics[width=\columnwidth]{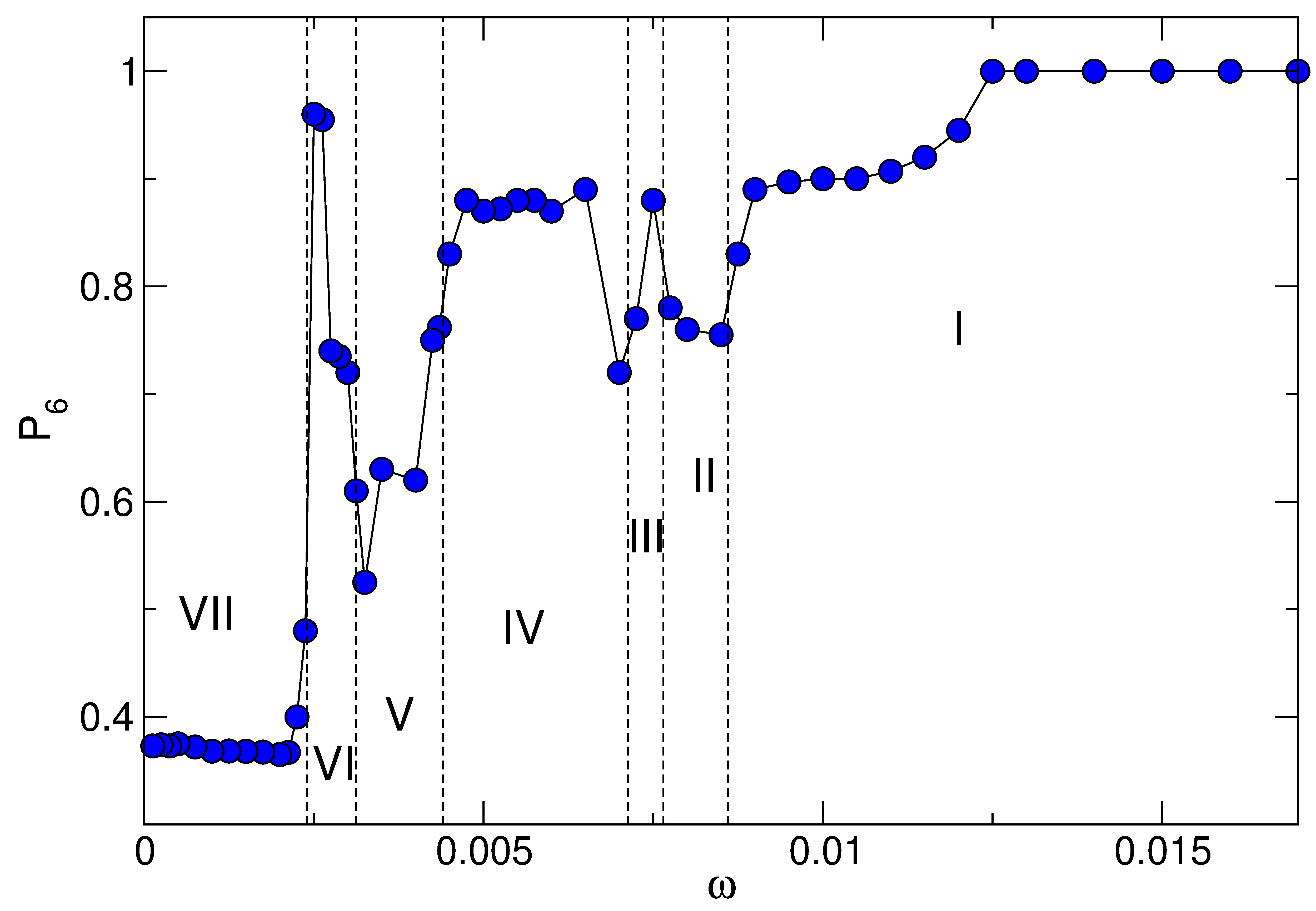}
\caption {The steady state value of $P_{6}$ vs $\omega$ for the
out of phase Bessel function particles at $\rho=0.208$ from
Fig.~\ref{fig:2}. Vertical lines separate
the different phases we identify, which are numbered from high to
low frequency.
I: distorted triangular crystal.
II: stripe lattice.
III: phase-separated solid.
IV: paired crystal.
V: phase-separated fluid.
VI: overlapping packed crystal.
VII: mixed fluid.
}
\label{fig:7}
\end{figure}

Although the Voronoi construction only takes into account the instantaneous
positions of the particles, and not their velocities,
it still captures the changes
that occur between states and can distinguish between ordered
and disordered states.
In Fig.~\ref{fig:7} we plot $P_{6}$ versus $\omega$
for the system from Fig.~\ref{fig:6},
where we take the average value of $P_6$
only after the system has reached a steady state.
The phases we distinguish are distorted triangular crystal (I),
stripe lattice (II), phase-separated solid (III), paired crystal (IV),
phase-separated fluid (V), overlapping packed crystal (VI), and
the mixed fluid state (VII).
For $\omega > 0.0086$, we find
a disordered triangular lattice (I) that
gradually becomes less distorted as $\omega$ increases.
Over the range $0.0086 < \omega < 0.012$,
the Voronoi algorithm gives an average value of
$P_6=0.9$ because the distortions of the triangular lattice are sufficiently
large that a finite number of non-sixfold coordinated polygons are present
in the system.
For $\omega > 0.012$, the triangular lattice is less distorted, all of the
Voronoi polygons are hexagons, and $P_6 = 1.0$.
When $0.0075 < \omega < 0.0086$, we observe a stripe lattice state (II) of
the type illustrated in
Fig.~\ref{fig:2}(c,d), and $P_6$ takes a lower value.
In the phase-separated solid (III) around $\omega=0.0075$,
$P_6$ increases to nearly $P_6\approx 0.9$
since for this frequency the system phase separates into only two
domains, while at a lower frequency of 
$\omega = 0.00725$, multiple phase separated domains are present,
which increases the amount of phase boundary in the system and consequently
depresses the value of $P_6$.
For $0.0045 < \omega < 0.0072$ in the paired crystal state (IV),
$P_6$ is close to $P_6 \approx 0.87$ but has
a dip near $\omega = 0.007$ due to the presence of
a small number of particles that do not form pairs.
In the phase-separated fluid (V),
$P_6$ is close to $P_6 \approx 0.6$,
which is
considerably higher than the
random mixed fluid (VII) value of $P_6 \approx 0.38$.
Between the phase-separated fluid and the random mixed fluid we find
a packed overlapping crystal (VI) with a maximum value of $P_6 \approx 0.95$,
but at higher values of $\omega$ in phase VI $P_6$ drops to
$P_6=0.75$, due to distortions of the Voronoi tesselation that occur
when particles get relatively close together.
There is also a small region from $0.002 < \omega < 0.0025$
in which the system is better described as a
partially phase-separated fluid, which is not highlighted in
Fig.~\ref{fig:7}.
For lower $\omega$, the system remains in
the mixed fluid state (VII) with $P_6 \approx 0.38$.

\begin{figure}
\includegraphics[width=\columnwidth]{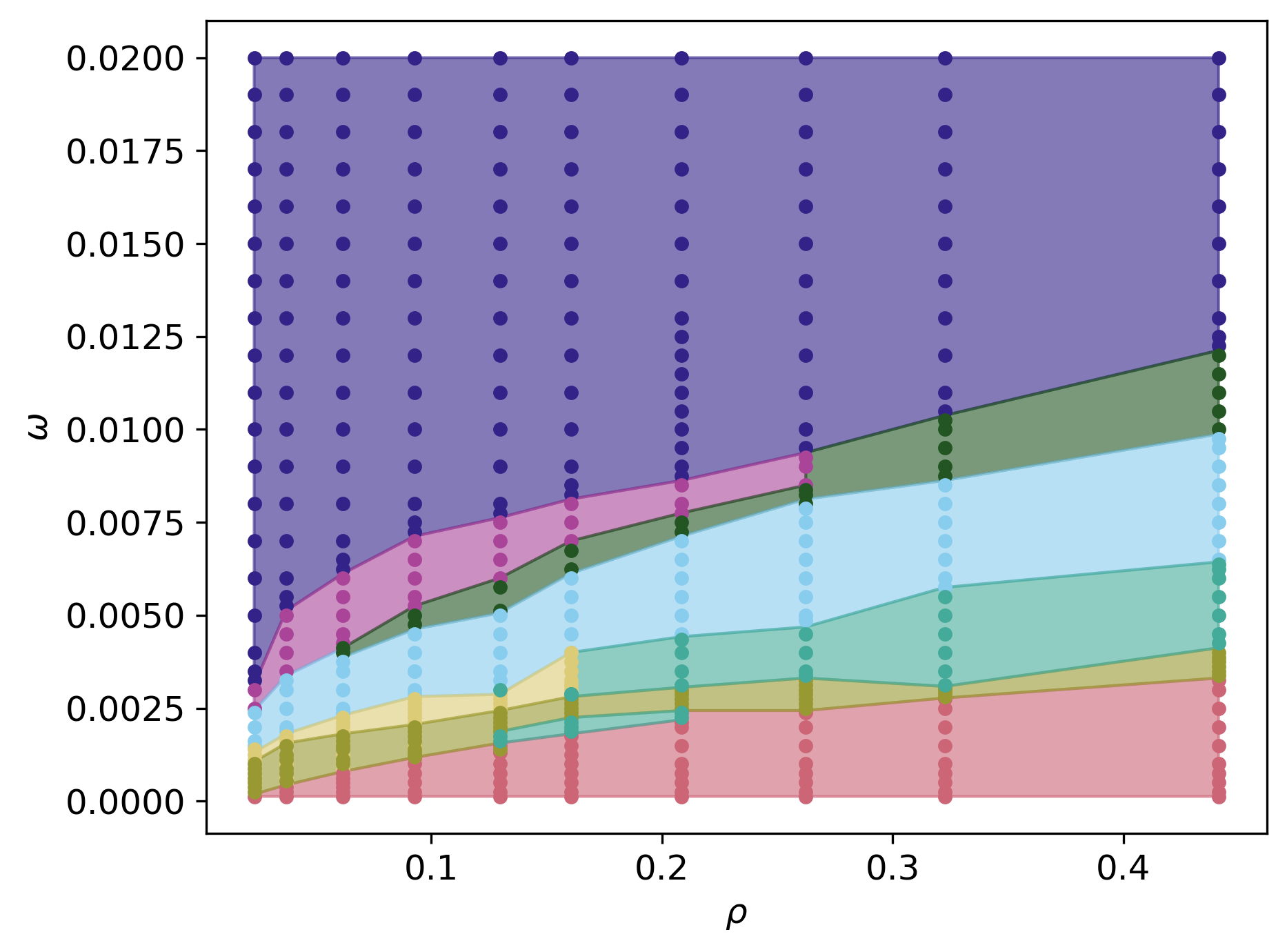}
\caption {Phase diagram for the out of phase Bessel function particles
from Fig.~\ref{fig:2} as a function of $\omega$ vs $\rho$.
Dark blue: distorted triangular crystal (I).
Light purple: stripe lattice (II).
Dark green: phase-separated solid (III).
Light blue: paired crystal (IV).
Light green: phase-separated fluid (V).
Dark yellow: overlapping packed crystal (VI).
Light yellow: overlapping stripe or glassy states (VI).
Red: mixed fluid (VII).
Example animations of motion in phases I, II, III, IV, VI, and VII
as well as in the noninteracting limit at $\rho=0.0926$
appear in the supplemental material \cite{Suppl}.
}
\label{fig:8}
\end{figure}

From the images of the particle motion and the behavior of $P_6$,
we use a range of particle densities to
construct the phase diagram shown in Fig.~\ref{fig:8}
as a function of $\omega$ versus $\rho$.
We highlight
the distorted triangular crystal (I),
stripe lattice (II),
phase-separated solid (III),
paired crystal (IV),
phase-separated fluid (V),
overlapping packed crystal (VI), and the
mixed fluid (VII).
For $\rho < 0.2$, we observe
overlapping stripe or glassy-like phases that we lump into
phase VI, 
and for lower $\omega$, there are
additional types of overlapping packed crystals.  
As $\rho$ increases, the lower boundary of the distorted triangular crystal
(I) shifts to higher $\omega$. This is due to the decrease in the
equilibrium lattice constant
which goes as $a \propto \rho^{-1/2}$,
so as $\rho$ becomes larger, it is necessary to apply
a higher $\omega$ to prevent the active orbits from overlapping.
For $\rho > 0.3$, the monomer stripe lattice state (II) disappears
due to the high elastic energy cost of elliptical orbits,
which increases with increasing $\rho$.
The paired crystal and overlapping crystal are present over the
entire range of densities shown in Fig.~\ref{fig:8},
and generally the phase boundaries shift to higher values of $\omega$ with
increasing $\rho$.
For lower $\rho$, the extent of the liquid phase is reduced
and a greater number of dynamically frozen states appear.
In principle, in the absence of thermal fluctuations for sufficiently
small $\rho$, the particles are so far apart that
they are only weakly interacting,
so the system should always settle into a frozen orbit state.

\begin{figure}
\includegraphics[width=\columnwidth]{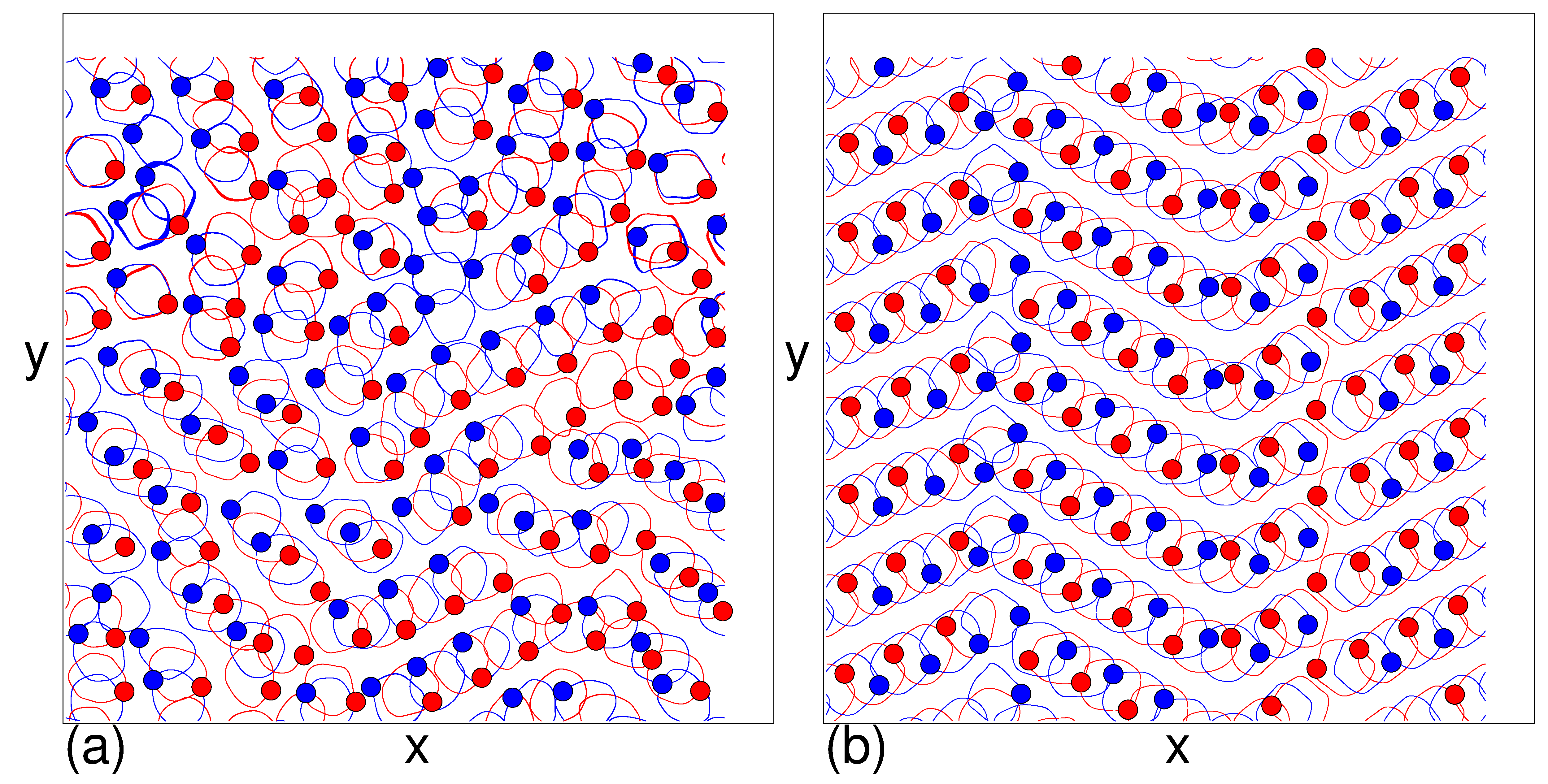}
\caption{Particle positions (circles) and trajectories (lines)
where species A is blue and species B is red for the out of phase Bessel
function particles at $\rho=0.161$ from the phase diagram in Fig.~\ref{fig:8}.
(a) A dynamical frozen stripe glass state (VI) at $\omega = 0.0035$.
(b) The system is more ordered at $\omega = 0.00325$.
}
\label{fig:9}
\end{figure}

In Fig.~\ref{fig:9}(a) we show an example of a dynamical
frozen stripe glass state at $\omega = 0.0035$ and $\rho = 0.161$.
In this case the paired particles overlap in chains, with the
particle species alternating
along the length of the chain.
In general, glassy states such as this one undergo a period of
transient rearrangements before settling into a state
where there is no long-time diffusion.
At these lower densities and frequencies,
the difference separating an ordered overlapping crystal from
a glassy state is a dynamical
commensuration effect
that determines which orbits keep the
moving particles as far apart from each other
for as much of the time as possible.
For example, at $\rho = 0.208$, when $\omega = 0.0055$
the individual particles are close enough together
that the ratio of the orbit radius to 
the equilibrium lattice spacing is 
$R_a/a \approx 1$, 
while at a frequency $\omega = 0.00275$ that is half as large,
$R_a/a \approx 2$. In each case, when $R_a/a$ is close to
an integer, an overlapping crystal appears.
The implication is that higher order overlapping lattices
could emerge whenever $R_a/a = n/m$ for integer $n$ and $m$,
and indeed, at the lower densities
we observe a variety of
overlapping crystal states.
When the commensurate conditions are not met,
the system forms either a fluid or a glassy state.
The interaction energy is higher in the glassy state than in the fluid,
and since
the interaction energy increases as
$\rho$ increases,
the fluid state is favored at higher densities and
we do not observe frozen glass states for $\rho > 0.2$.
Frozen glasses do appear
at lower $\rho$ when the particle-particle spacing is larger.
In some cases, the stripe glasses exhibit a noticeable remnant of
ordering, as shown in Fig.~\ref{fig:9}(b)
at a lower frequency of $\omega = 0.00325$.
The ordering in the stripe glass or other overlapping packed crystal
states could also be affected by the
periodic boundary conditions of our system.
Geometric confinement could produce
other types of crystal states,
and the system could be fluid-like along the
edges of the confinement but crystalline in the bulk.
There is a transient time that must elapse before a glassy system
reaches a frozen state, and the length of this transient is expected to
increase with increasing system size, so it is possible that in a
sufficiently large system
the glassy arrangement would never reach
a truly frozen state on observable time scales.
We find, however,
that the transient motion in the glassy state
occurs in bursts and that the system spends some time frozen before
undergoing
intermittent rearrangements, a dynamical process that will be
explored elsewhere.

\begin{figure}
\includegraphics[width=\columnwidth]{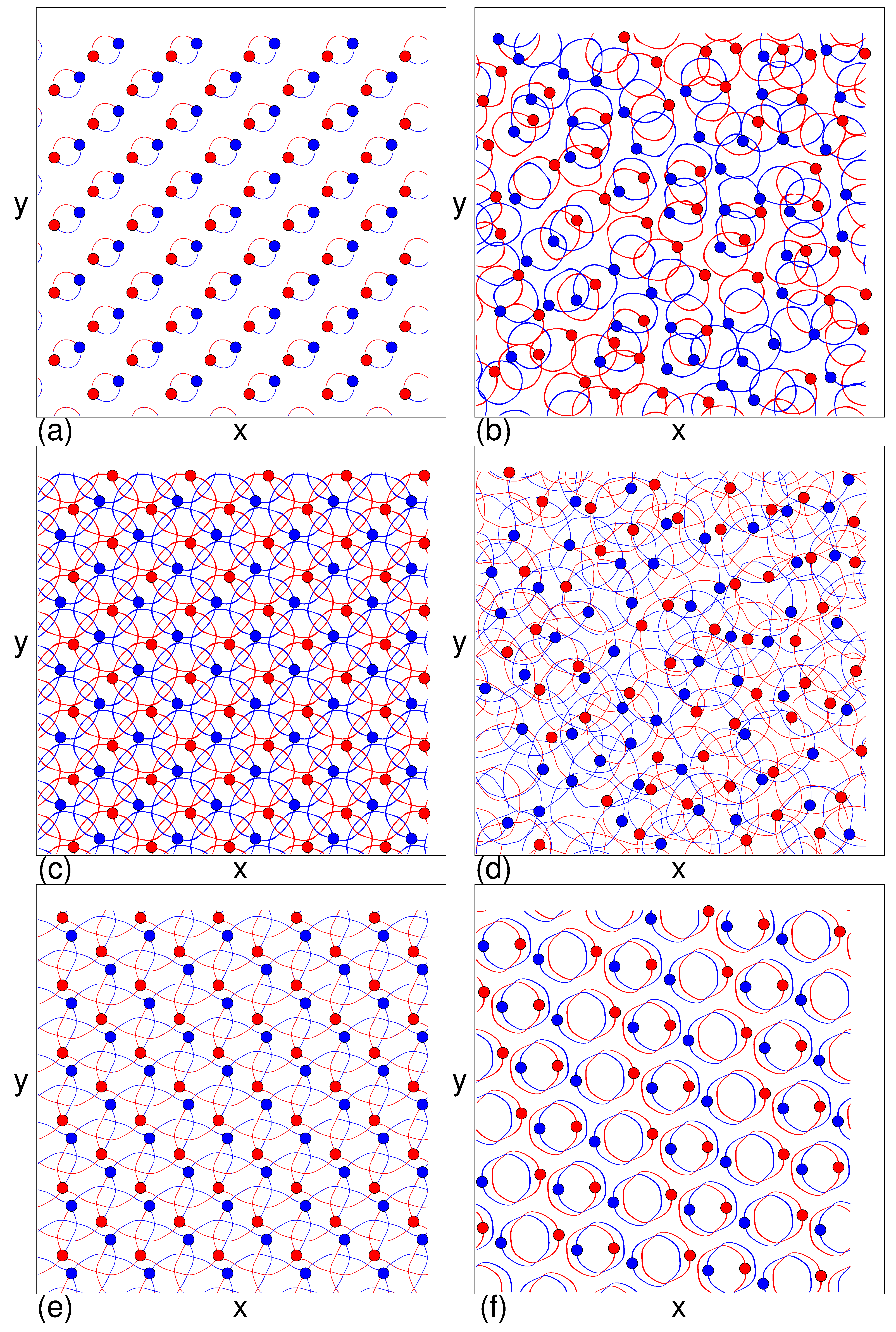}
\caption{Particle positions (circles) and trajectories (lines)
where species A is blue and species B is red for the out of phase Bessel
function particles
from Fig.~\ref{fig:8}.
(a) Paired crystal (IV) at $\rho = 0.0926$ and $\omega = 0.004$.
(b) Stripe glass at $\rho = 0.0926$ and $\omega = 0.0025$.
(c) Overlapping packed crystal (VI) at
$\rho = 0.0926$ and $\omega = 0.00175$.
(d) Disordered glass at $\rho = 0.0926$ and $\omega = 0.0015$.
(e) Overlapping packed crystal (VI) at $\rho = 0.0926$ and $\omega = 0.00125$.
(f) Dimer lattice at $\rho = 0.0617$ and $\omega = 0.00225$.
}
\label{fig:10}
\end{figure}

In Fig.~\ref{fig:10}(a) we show the particle configurations and trajectories in
the paired crystal (IV) state
for the system from
Fig.~\ref{fig:8} at $\rho = 0.0926$ and $\omega = 0.004$.
Figure~\ref{fig:10}(b) shows that for 
the same density at $\omega = 0.0025$, the system organizes into a
stripe glass, while in
Fig.~\ref{fig:10}(c) at $\omega = 0.00175$,
there is an overlapping packed crystal (VI).
In Fig.~\ref{fig:10}(d) at $\omega = 0.0015$,
the system freezes into a disordered glass state.
When $\omega = 0.00125$, as in
Fig.~\ref{fig:10}(e),
another overlapping packed crystal (VI)
appears that is different from the one
found at $\omega = 0.00175$.
There are also
other ordered stripes and even dimer states
that can form,
such as the state
shown in Fig.~\ref{fig:10}(f) at $\rho = 0.0617$ and $\omega = 0.00225$
which can be described as a dimer lattice.
At smaller densities,
the periodic boundary conditions can
start to affect the ordering;
however, when we consider
the same sets of parameters in
larger systems with $L = 48$ and $L=72$,
we generally obtain the same sets of phases. In some
cases the stripe glass states can be more ordered or disordered
when the system is made larger, but the other ordered phases
remain robust.

\section{Screened Coulomb Interactions}

\begin{figure}
\includegraphics[width=\columnwidth]{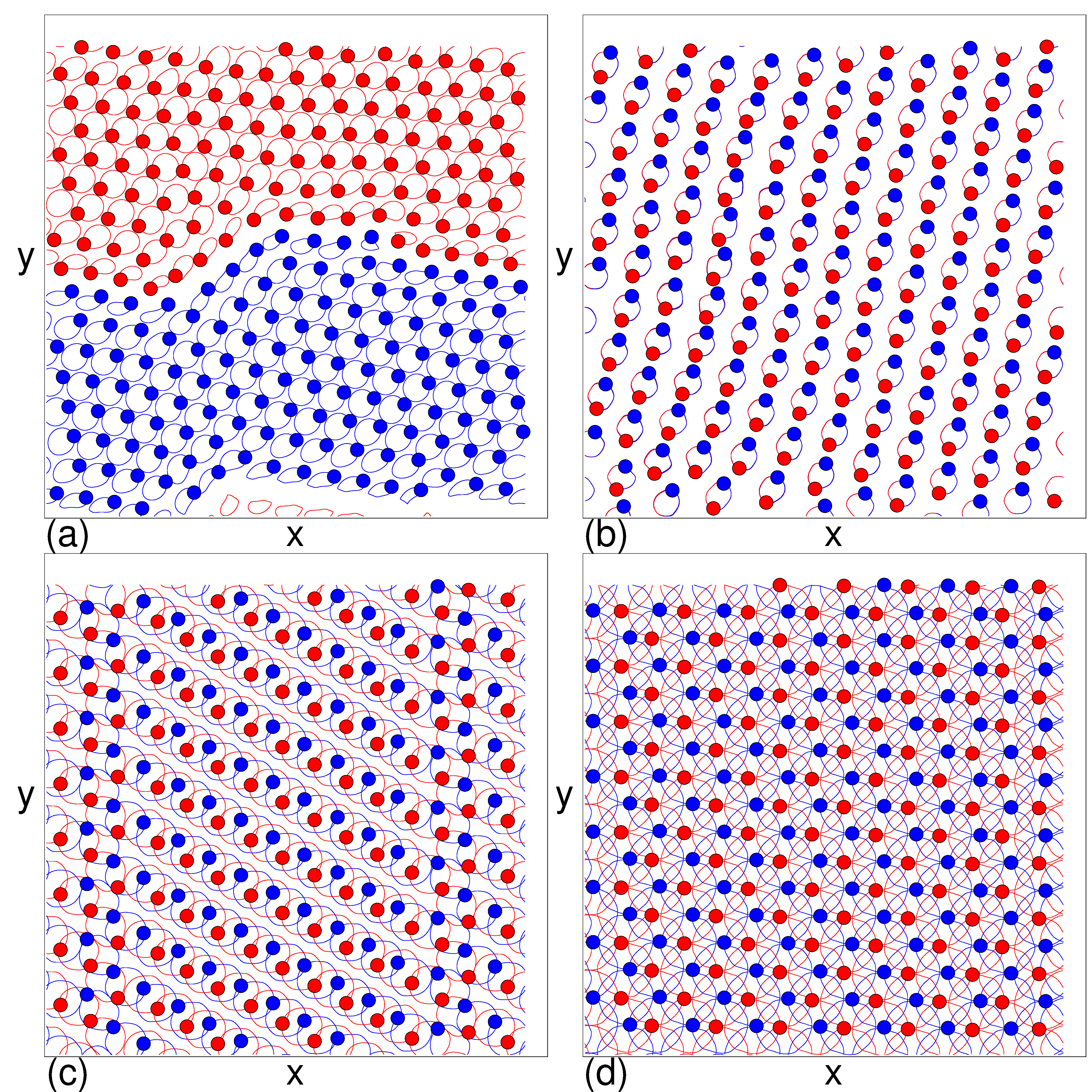}
\caption{Particle positions (circles) and trajectories (lines)
where species A is blue and species B is red for the out of phase  
Yukawa particles
with interactions of the form $V(r) = C\exp(-r)/r$
at $\rho = 0.208$.
(a) A phase-separated solid (III) at $C = 0.5$ and $\omega=0.00525$.
(b) A paired crystal (IV) at $C = 0.2$ and $\omega=0.00525$.
(c) A frozen disordered state at $C=0.02$ and $\omega = 0.035$.
(d) An overlapping packed crystal (VI)
at $C=0.02$ and $\omega = 0.025$.
}
\label{fig:11}
\end{figure}

We next consider the case of particles with
screened Coulomb or Yukawa interactions.
We observe a similar set of phases
as for the Bessel function interactions;
however, the coefficient in the interaction potential must take
different values in the two systems to give the same phases.
For $V(r) = C\exp(-r)/r$,
if we take $\rho = 0.208$ and $\omega = 0.00525$,
where we found a paired crystal (IV)
for the Bessel function interaction,
we instead find a disordered phase-separated state (III) at $C=0.5$,
as shown in Fig.~\ref{fig:11}(a).
In this state, the particles are able to form closed circular orbits, but
diffusion can still occur.
For $C = 0.2$, we find a paired crystal (IV)
as shown in Fig.~\ref{fig:11}(b).
If $C$ is too small, the system forms a disordered frozen state,
as shown at $\omega=0.035$ and $C=0.02$ in Fig.~\ref{fig:11}(c).
For the same low $C=0.02$ at $\omega=0.025$ in
Fig.~\ref{fig:11}(d),
there is an overlapping packed crystal (VI).
In general, for smaller $C$, the fluid phase is reduced
in extent and a greater variety of
overlapping packed crystal states can form.
If we consider larger values of $C$, many of the same phases are still present
but have shifted 
to different densities.
These results indicate 
that the phases we observe should be robust
over a wide range of intermediate repulsive particle-particle interaction
potentials.

\section{Counter-Rotating Mixtures}

\begin{figure}
\includegraphics[width=\columnwidth]{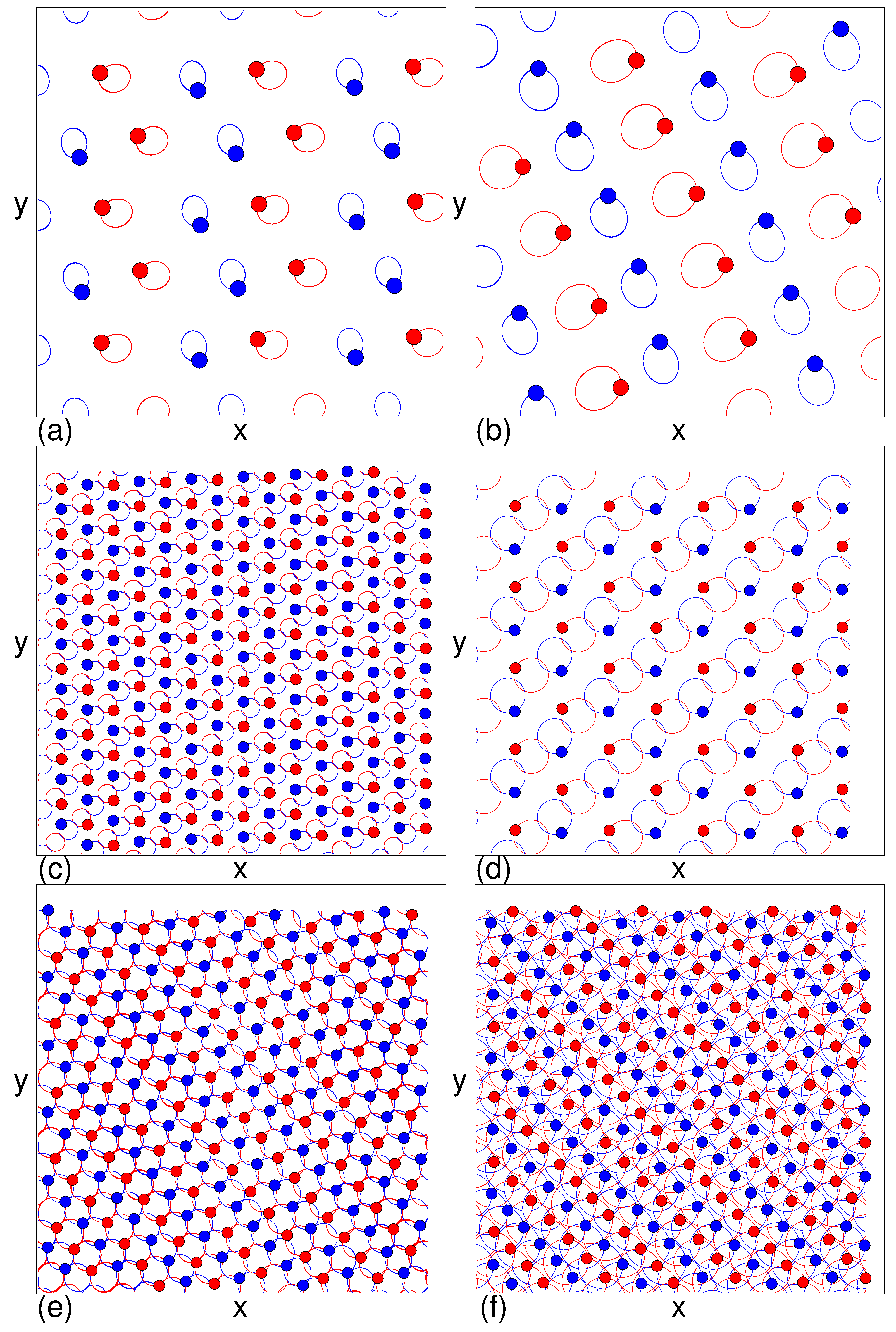}
\caption{Particle positions (circles) and trajectories (lines)
where species A is blue and species B is red for the counter-rotating
Bessel function particles shown in Fig.~\ref{fig:1}(c).  
(a) A blow-up view of the distorted triangular crystal (I) at
$\rho = 0.208$ and $\omega = 0.011$.
(b) A blow-up view of the
stripe lattice (II) at $\rho = 0.208$ and $\omega = 0.008$.
(c) An ordered chain state at $\rho = 0.208$ and $\omega = 0.006$.
(d) An ordered chain state at $\rho=0.0617$ and $\omega = 0.0035$.
(e) An overlapping lattice with some dislocations
at $\rho = 0.208$ and $\omega = 0.004$.
(f) An overlapping packed crystal (VI) at $\rho = 0.208$ and
$\omega = 0.0025$.
}
\label{fig:12}
\end{figure}

We return to the Bessel function interactions and
next consider a counter-rotating system in which one particle species
rotates in one direction and the other
species rotates in the opposite direction, as shown in
Fig.~\ref{fig:1}(c).
Although
some of the dynamical phases we find are similar to those from the
system where the two particle species are rotating in the same direction but
out of phase,
there are numerous differences.
We use the Bessel function interaction to make a clear comparison with
the out of phase system from Fig.~\ref{fig:2}.
In Fig.~\ref{fig:12}(a) we show the particle locations and trajectories
for a counter-rotating system
at $\rho = 0.208$ and $\omega = 0.011$,
where we highlight only a subset of the system for clarity.
We find a distorted triangular lattice (I)
where the particles have non-overlapping
elliptical orbits and the ellipses of each species are tilted in a
different direction.
For these parameters, the out of phase system forms
a similar state.
At high frequencies,
the orbits of adjacent particles
are not large enough to overlap,
so the pairwise repulsion between the particles dominates, producing
a similar triangular lattice regardless of whether the particles are
rotating in the same direction out of phase or are counter-rotating.

In Fig.~\ref{fig:12}(b), at $\omega = 0.008$ the counter-rotating
system forms a non-overlapping stripe-like lattice
with larger orbits that are still elliptical with two tilt directions.
When $\omega=0.006$, where a paired crystal (IV) appears
for the out of phase system, Fig.~\ref{fig:12}(c)
shows that the counter-rotating system
forms a periodic array of chains with overlapping orbits,
indicating that the different species have
effective dipolar or anisotropic interactions with each other.
Formation of a chain structure permits
the particles of opposite chirality to
move in such a way that two neighboring particles never approach
each other closely
throughout the cycle.
The pairwise repulsion also produces
a well-defined spacing between adjacent chains.
In general, we find that for parameters at which the out of phase
particles are in a paired crystal state (IV),
there is an overlapping stripe state for counter-rotating particles.
We show another example of such an overlapping stripe state
in Fig.~\ref{fig:12}(d)
at $\omega = 0.0035$ and $\rho = 0.0617$.
It is easy to understand why the counter-rotating particles do not form
a paired crystal,
since if particles with opposite chirality were
to share the same orbit, they would collide and break apart the orbit.
The counter-rotating particles exhibit some phases that do not appear
in the out of phase system. For example,
in Fig.~\ref{fig:12}(e) for $\rho = 0.208$ and $\omega = 0.004$,
we find an overlapping lattice that contains some dislocations.
For the same system at $\omega=0.0025$ in Fig.~\ref{fig:12}(f),
an ordered overlapping packed crystal (VI) appears.
In general, there are a greater amount of ordered dynamical states for
the counter-rotating particle mixtures than for the out of phase particles,
since the opposite chirality allows the particles to remain further apart
throughout the driving cycle compared to
when the particles have the same chirality but are out of phase.
It would also be possible to explore
other driving protocols in which the phase difference
between the particles of the same chirality
is not $180^\circ$ but has some other value, and this could
produce an even richer set of phases.

\begin{figure}
\includegraphics[width=\columnwidth]{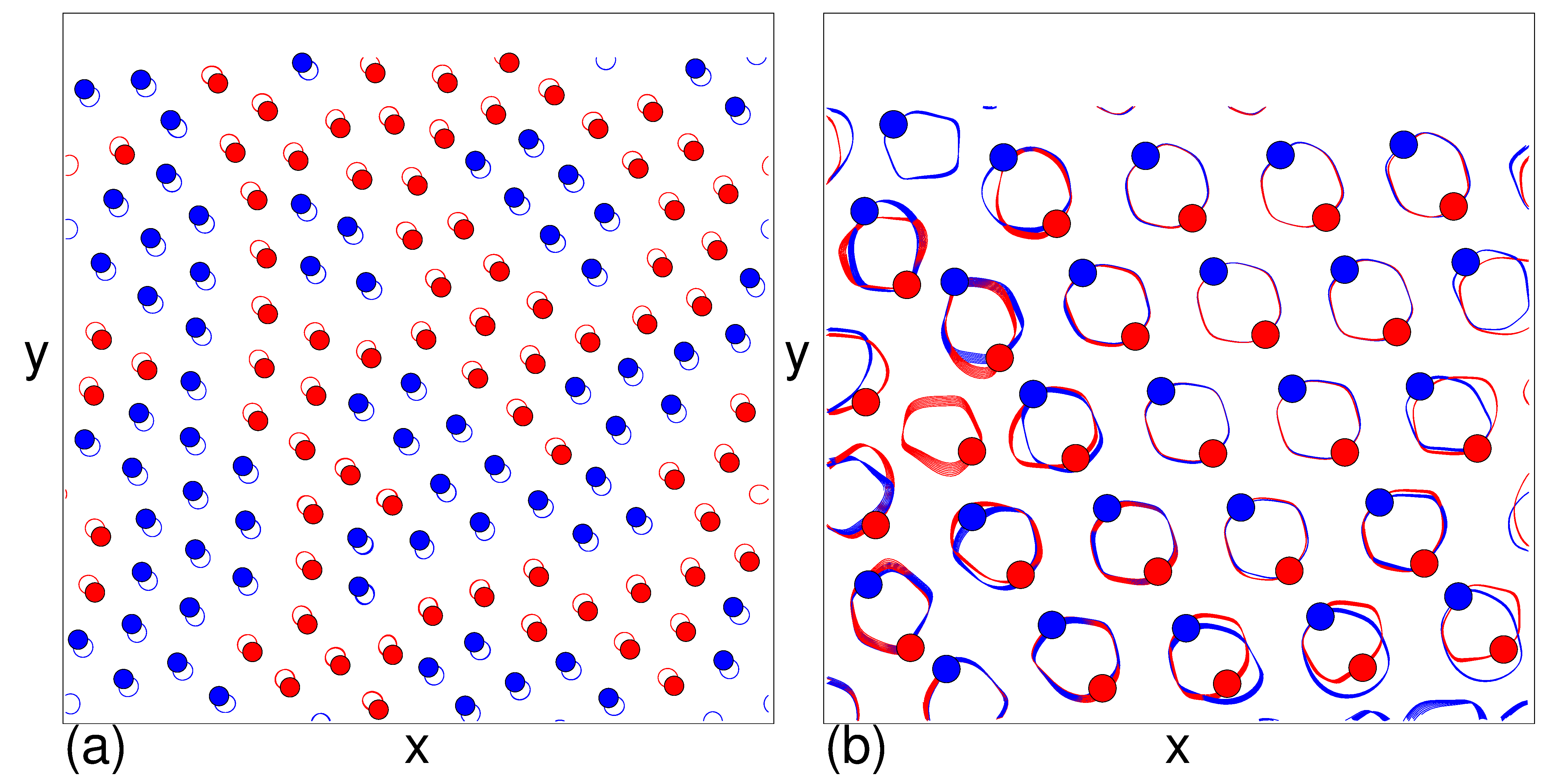}
\caption{Particle positions (circles) and trajectories (lines)
where species A is blue and species B is red for the out of phase Bessel
function particles from Fig.~\ref{fig:1}(b) where the initialization
protocol has been changed and the particles are placed at random
non-overlapping positions.  
The images are obtained in the steady state
after the system has passed through a
transient disordered triangular lattice.
(a) At $\rho = 0.208$ and $\omega = 0.01$,
we find a distorted and partially phase-separated triangular lattice.
(b) At $\rho = 0.208$ and $\omega = 0.00525$, a zoomed in view shows that
the system forms a paired crystal containing some unpaired particle
defects.
}
\label{fig:13}
\end{figure}

\section{Initial Preparation}

Up to this point we have initialized the system in a
triangular lattice where each species occupies every other lattice site in an
ordered tiling. To test the effect of the initialization protocol on the
results, we place the particles at randomly chosen non-overlapping positions,
with the same 50:50 distribution of species A and B but with no ordering in
the starting locations of each species.
At high $\omega$, we still obtain a distorted triangular crystal (I) as in
the phase diagram of Fig.~\ref{fig:8}, but the arrangement of the species in the
crystal is more mixed. Species A and B are randomly distributed inside the
crystal state instead of forming wavy vertical lines as in Fig.~\ref{fig:2}(a).
For these high frequencies, the orbits are sufficiently small that the
system behaves like a collection of point particles that are insensitive
to the chirality, so the system simply crystallizes without regard to particle
species.
In Fig.~\ref{fig:13}(a), we show the particle
configurations and trajectories
for an assembly of out of phase Bessel function particles
at $\rho = 0.208$ and $\omega = 0.01$, where we find
a distorted triangular lattice that has some amount of
phase separation.
The stripe crystal of Fig.~\ref{fig:2}(b)
has been replaced by a disordered mixed triangular lattice state due to the
change in the initialization protocol.
The two states appear over nearly the same
sets of parameters for both initialization protocols.
For values of $\rho$ and $\omega$ where the paired crystal occurs
in Fig.~\ref{fig:8}, the
randomly initialized system still forms pairs,
but there are a small number of particles that are unable to find a pair,
resulting in the emergence of trapped monomers.
There is also a longer transient time for the randomly initialized system
before it settles in to the
mostly paired state.
In Fig.~\ref{fig:13}(b),
we show the particle configuration and trajectories at
$\rho = 0.208$ and $\omega = 0.00525$ for the randomly initialized system
after it has settled into a
steady state. Most of the particles are paired but there are a few
monomer defects present.
At lower frequencies and densities,
changing the initialization protocol to random gives
stripe glass states and overlapping crystal states
that are more disordered compared to the triangular lattice initialization
protocol, but in general, at the lowest densities,
the system still
reaches a dynamical frozen state.

\section{Particles with Coulomb Interactions}

\begin{figure}
\includegraphics[width=\columnwidth]{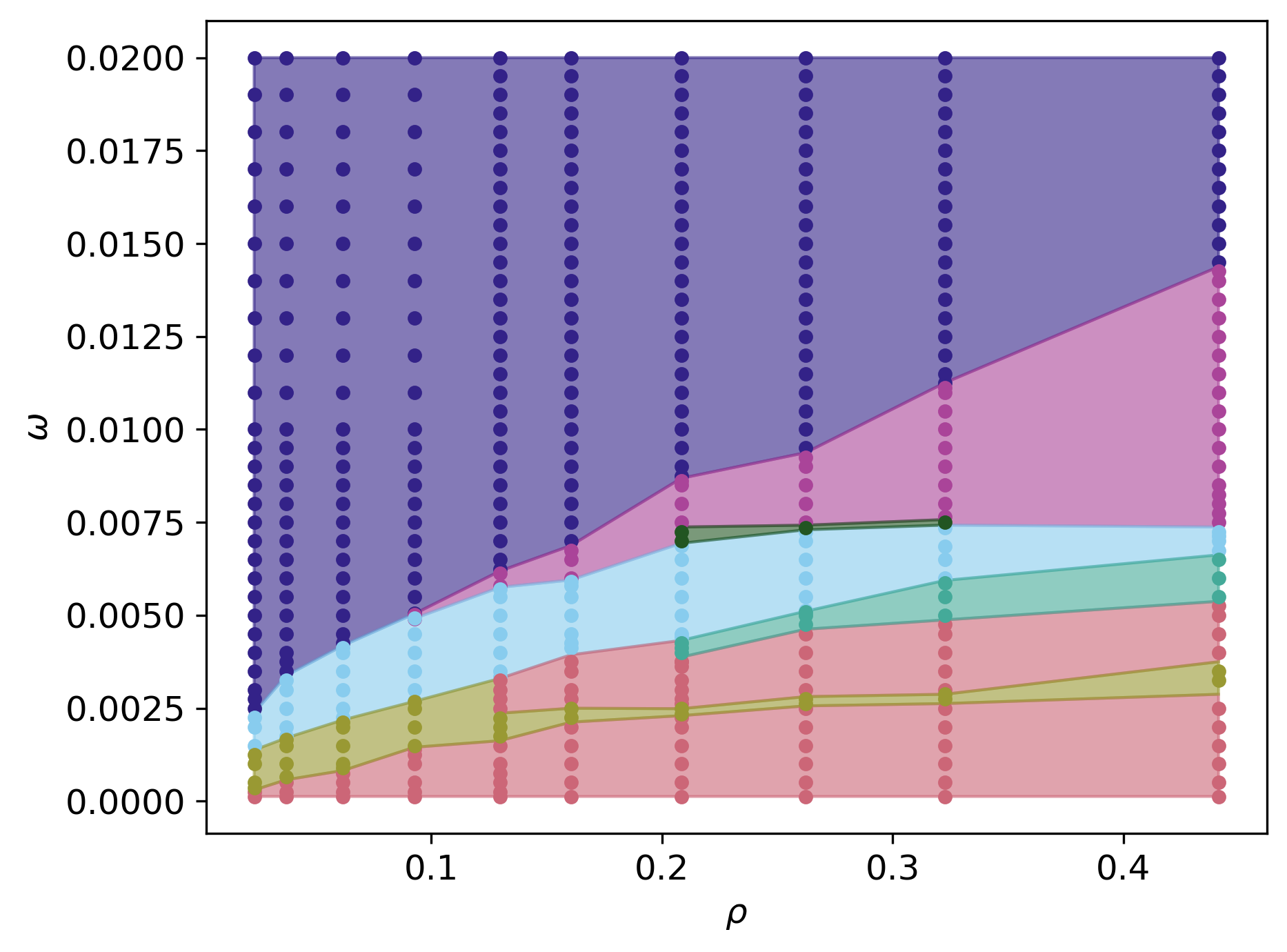}
\caption {Phase diagram for the out of phase Coulomb particles
as a function of $\omega$ vs $\rho$.
Dark blue: distorted triangular crystal (I).
Light purple: stripe lattice (II).
Dark green: phase-separated solid (III).
Light blue: paired crystal (IV).
Light green: phase-separated fluid (V).
Dark yellow: overlapping packed crystal and overlapping stripe or glassy
states (VI).
Red: mixed fluid (VII).
}
\label{fig:14}
\end{figure}

We return to our original initialization protocol and consider the effect
of using 
particles with long-range Coulomb interactions, beginning with particles
that have the same chirality but that are driven out of phase by $180^\circ$.
We find that the dynamics are similar to those of the shorter-range
repulsively interacting systems with Bessel or Yukawa interactions.
In Fig.~\ref{fig:14}, we show the phase diagram for the Coulomb system
as a function of $\omega$ vs $\rho$, where we find the same phases from
before: distorted triangular crystal (I), stripe lattice (II), phase-separated
solid (III), paired crystal (IV), overlapping packed crystal and overlapping
phase-separated fluid (V),
stripe or glassy states (VI),
and mixed fluid (VII).
The overall shape of the phase diagram is similar to what we find for the
intermediate-range repulsive Bessel function particles in Fig.~\ref{fig:8}.
In general, the phase-separated states are less prominent for the Coulomb
particles than for the Bessel particles. Note that
in Fig.~\ref{fig:14}, we do not distinguish between the ordered
overlapping packed crystal and the stripe glass states.

\begin{figure}
\includegraphics[width=\columnwidth]{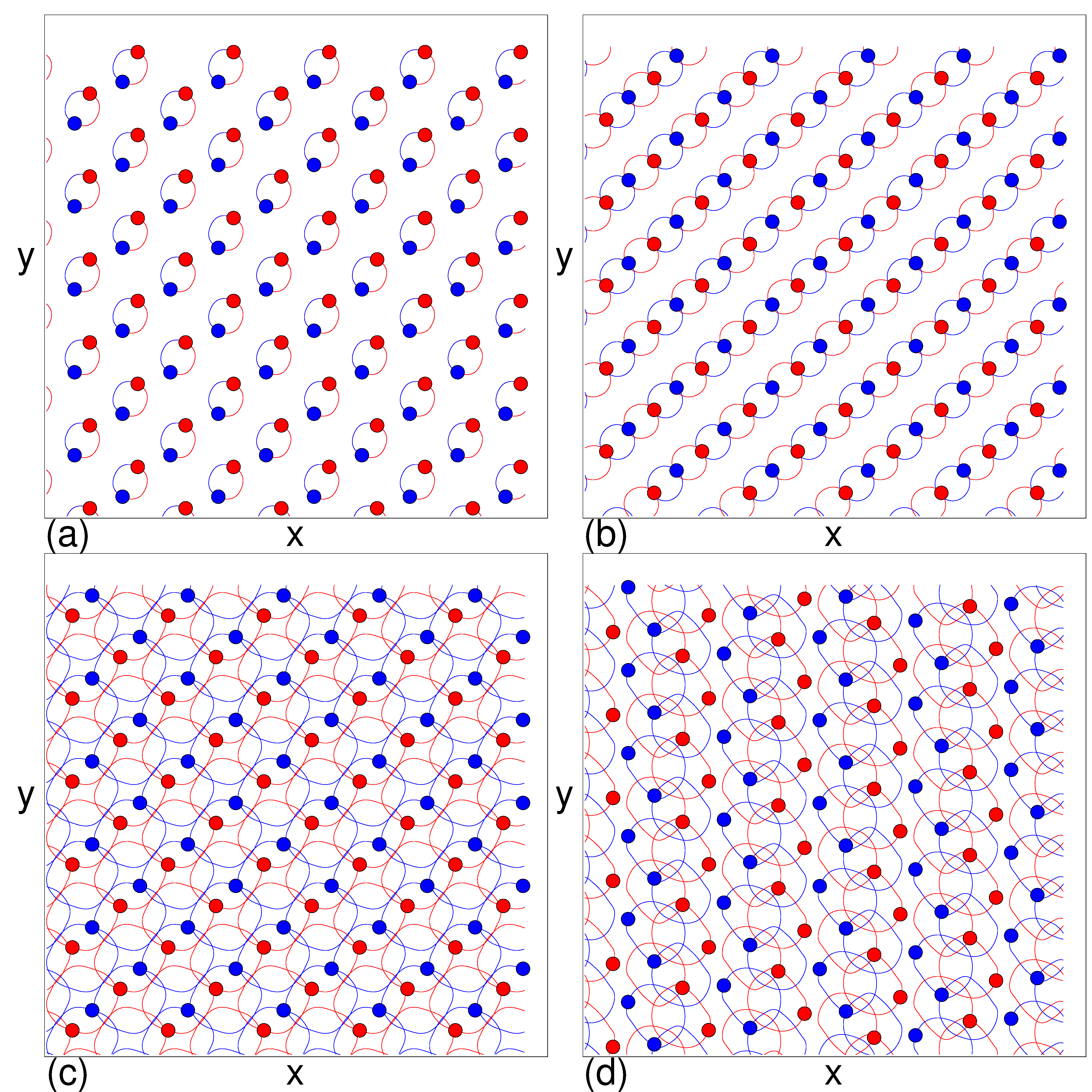}
\caption{Particle positions (circles) and trajectories (lines)
where species A is blue and species B is red for the 
Coulomb particles from the system in Fig.~\ref{fig:14}.
(a) Paired crystal (IV) at $\rho = 0.0923$ and $\omega= 0.004$
for out of phase particles.
(b) A chain lattice state at the same $\rho = 0.0923$ and $\omega= 0.004$ as
in panel (a) but for counter-rotating particles.
(c) Overlapping packed crystal at
$\rho = 0.0923$ and $\omega= 0.002$ for out of phase particles.
(d) Overlapping stripe crystal at
$\rho = 0.0923$ and $\omega= 0.00225$ for out of phase particles.}
\label{fig:15}
\end{figure}

In Fig.~\ref{fig:15}(a) we show the particle positions and trajectories
for the out of phase Coulomb particles from
Fig.~\ref{fig:14} at $\rho = 0.0923$ and $\omega= 0.004$.
Here there is a paired crystal lattice (IV) that has the same features as
the paired crystals observed for Bessel and Yukawa interacting particles.
At the same density and frequency of $\rho=0.0923$ and $\omega=0.004$, if
the particles instead counter-rotate as in Fig.~\ref{fig:1}(c),
Fig.~\ref{fig:15}(b) shows
that an ordered chain state appears.
In general we do not observe paired crystal states
for counter-rotating particles with opposite chirality for the
Coulomb system.
For the out of phase Coulomb particles from Fig.~\ref{fig:14},
Fig.~\ref{fig:15}(c) illustrates
the overlapping packed crystal (VI)
that appears at $\rho = 0.0923$ at $\omega= 0.002$,
while in Fig.~\ref{fig:15}(d), we find
an overlapping stripe crystal (VI) at $\rho = 0.0923$ and $\omega= 0.00225$
in the same system.

\begin{figure}
\includegraphics[width=\columnwidth]{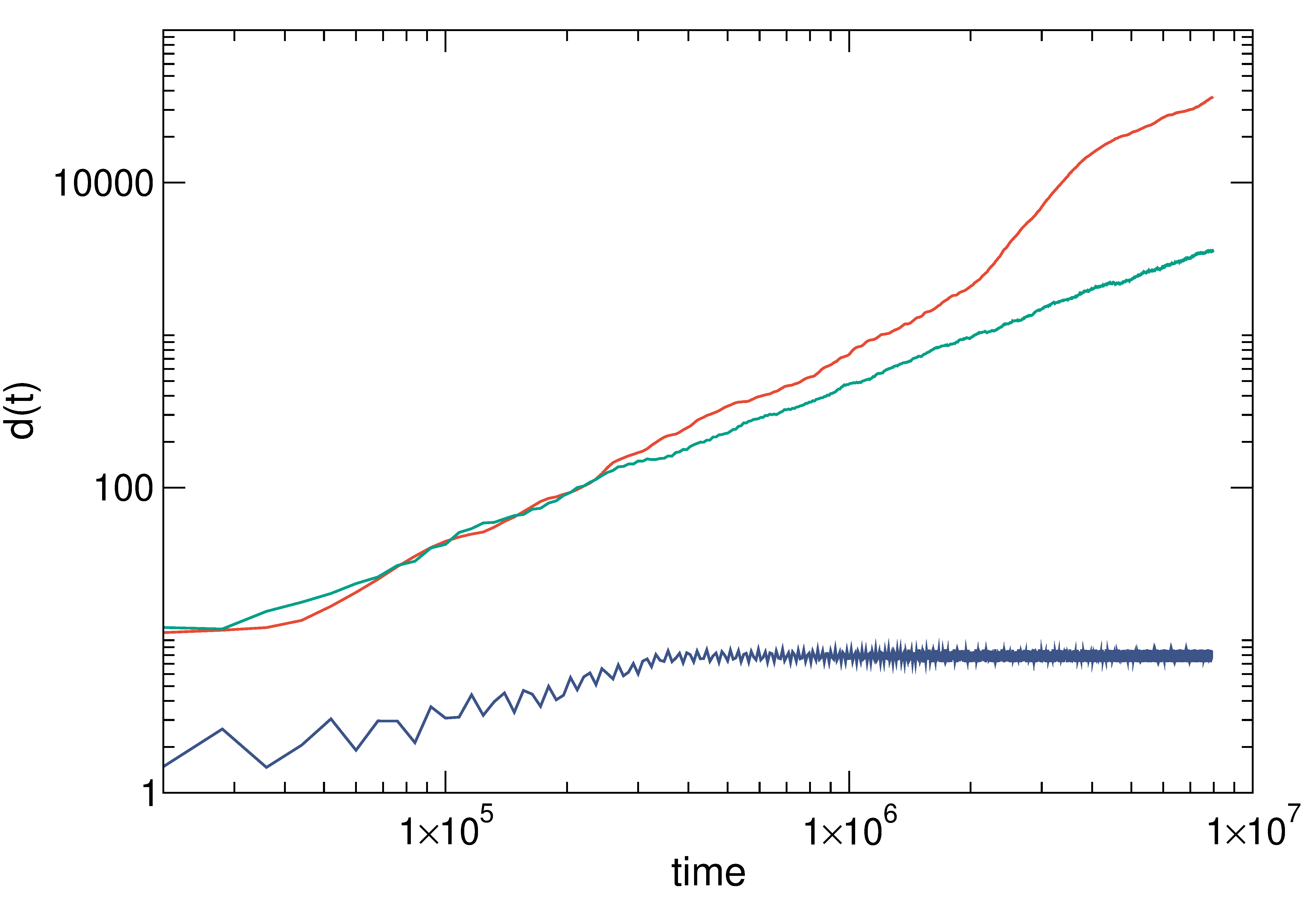}
\caption{The root mean square displacement $d(t)$
vs $t$ for the out of phase Coulomb system from Fig.~\ref{fig:14}
at $\rho = 0.208$.
Blue: the paired crystal (IV) state at
at $\omega = 0.005$.
Red: the phase-separated fluid (V) at $\omega=0.004$.
Green: the mixed fluid (VII) at $\omega = 0.002$.
}
\label{fig:16}
\end{figure}

Another method for categorizing the different phases is by measuring
the mean square displacement $d(t)$ versus time.
In general, for crystal phases,
$d(t)$ shows a
short-time transient behavior followed by saturation to a constant
value, since
there is no long-time diffusion.
In Fig.~\ref{fig:16}, we plot $d(t)$
for the out of phase Coulomb system
from Fig.~\ref{fig:14} at $\rho = 0.208$.
For the paired crystal state (IV) at $\omega = 0.005$, $d(t)$
is flat at long times. We observe similar behavior in
the overlapping packed crystal and frozen glass phases since none of these
states has any
long-time diffusion in the frozen regime.
The transient time required
to reach a disordered frozen glass state is
generally longer than the time needed to form a crystal state.
For the mixed fluid (VII) at $\omega = 0.002$,
$d(t)$ increases linearly with time,
indicating that regular diffusion is occurring.
The phase-separated fluid at $\omega = 0.004$
exhibits diffusive motion at early times and a more rapid
increase in $d(t)$ at intermediate times as the system phase separates. This
is followed
by regular diffusion at long times in the phase-separated state.
Along the boundary between the domains of different particle species in
the phase-separated fluid, there is edge transport
where particles can temporarily move ballistically;
particles can also move in and out of the boundary.
For the phase-separated solid (III, not shown), the form of
$d(t)$ is similar to that found for
the paired crystal (IV),
where $d(t)$ reaches a plateau value at long times since
there is no long-time diffusion.

\section{Thermal Fluctuations and Melting in the Paired Crystal State}

We next address the question of how stable the phases we observe are
against the introduction of thermal fluctuations.
In general we find that the phases are robust in the presence of
a finite temperature,
and that interesting melting behaviors occur
that will be explored in another work.
Here we focus on
the paired crystal state,
which is not only stable and robust against thermal fluctuations,
but also can persist
up to temperatures
that are higher than the melting temperature $T_m$ of the
nondriven lattice.

\begin{figure}
\includegraphics[width=\columnwidth]{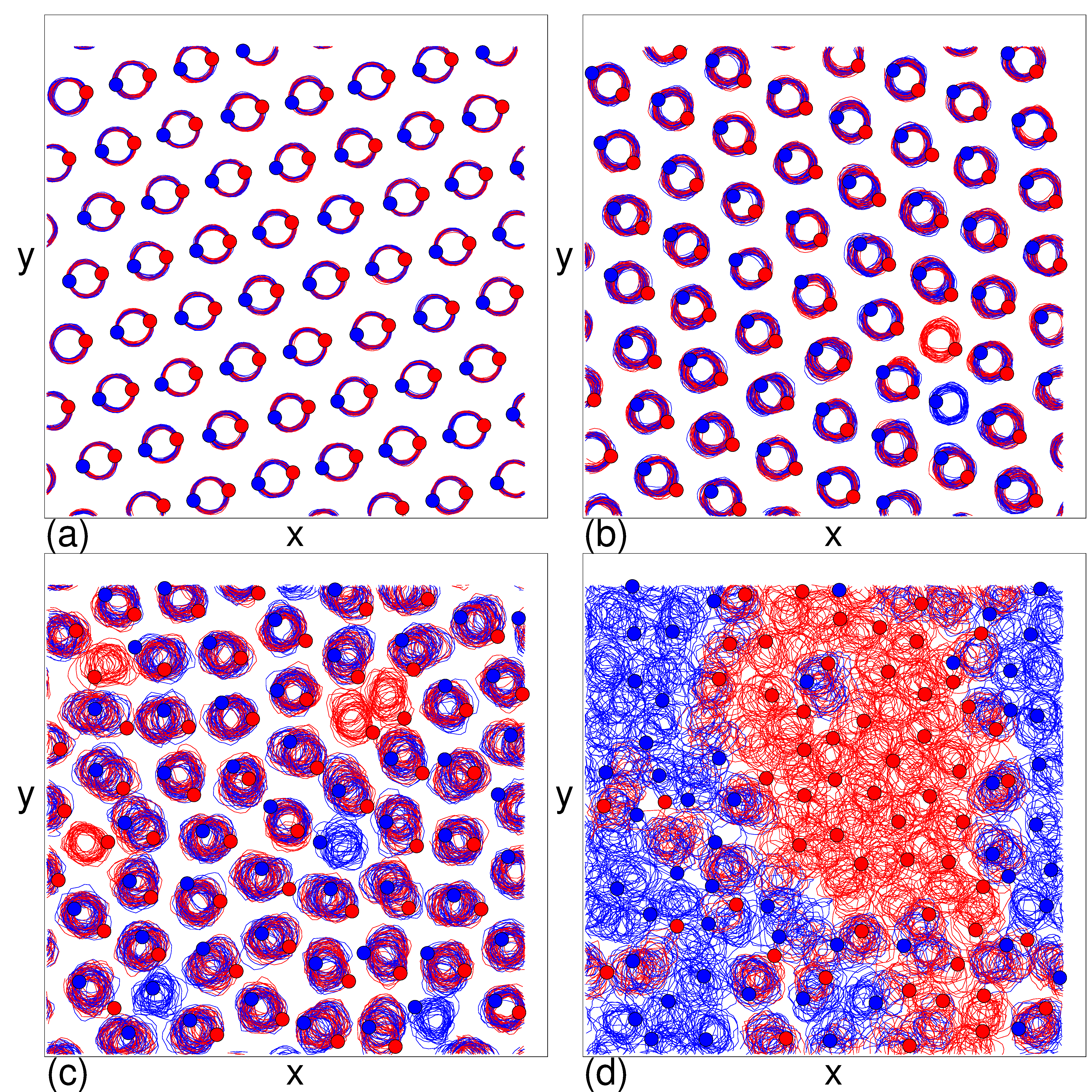}
\caption {Particle positions (circles) and trajectories (lines)
where species A is blue and species B is red
for the out of phase Coulomb particles from Fig.~\ref{fig:14}
at $\rho = 0.0923$ and $\omega = 0.004$,
where a paired crystal (IV) forms at zero temperatures.
We examine thermal effects at different values of
$T/T_m$, where $T_m$ is the temperature at which the
non-driven system first shows long-time diffusion.
(a) At $T/T_m = 0.33$ there is a paired crystal.
(b) At $T/T_m = 0.67$ there is also a paired crystal.
(c) At $T/T_m = 1.33$ a paired fluid appears.
(d) At $T/T_m = 2.0$ there is a phase-separated fluid.
}
\label{fig:17}
\end{figure}

We represent the thermal forcing with
Langevin kicks that have
the properties
$\langle F^{T}(t)\rangle = 0.0$
and $\langle F^T_i(t)F_j^T(t^\prime)\rangle = 2\eta k_BT\delta_{ij}\delta(t- t^\prime)$, and
we consider the out of phase Coulomb interacting system
in the paired crystal state (IV) at $\rho = 0.09529$ and $\omega = 0.004$.
When the particles are passive and the ac driving is zero,
the system forms a triangular lattice in which
the onset of diffusion occurs at a temperature $T_m$
which we define as the equilibrium melting temperature.
When we make the system active by turning on the ac driving,
the paired crystal remains robust up to temperatures that are
slightly higher than $T_m$.
In Fig.~\ref{fig:17}(a), we show the
particle positions and trajectories
at $T/T_m = 0.33$.
The orbits show some thermal smearing,
but the pairs are persistent and the paired crystal lattice remains intact.
At $T/T_m=0.67$ in Fig.~\ref{fig:17}(b),
the thermal smearing has widened and a single unpaired defect has appeared,
but apart from this highly localized distortion
the crystal structure remains present and there is no long-time diffusion.
Figure~\ref{fig:17}(c) shows the motion at $T/T_m = 1.33$,
where long-time diffusion is occurring. Most of the particles remain in
bound pairs, so the system is best described as a paired fluid.
This result indicates that the binding energy for the pairing of the particles
is larger than the elastic energy of the paired crystal lattice.
For $T/T_m=2.0$ in Fig.~\ref{fig:17}(d), most of
the pairs have broken apart and
strong diffusion occurs throughout the system.
Some phase separation appears,
and we find
some transient pairing along the boundaries of the
phase domains.
At higher temperatures than what is illustrated here,
the system forms a mixed fluid.
A number of open questions would be interesting to explore in
future work. These include whether
the paired crystal to paired fluid transition is associated with a
paired hexatic state, such as the hexatic phases
recently observed
in charged colloidal systems with additional activity \cite{Vyas26}.
It would also be interesting to determine
if there is multiple-step melting, as well as
the transport properties of the paired fluid,
how the other phases melt, and how the counter-rotating
system melts.

\section{Phase Time Molecular Crystals}

\begin{figure}
\includegraphics[width=\columnwidth]{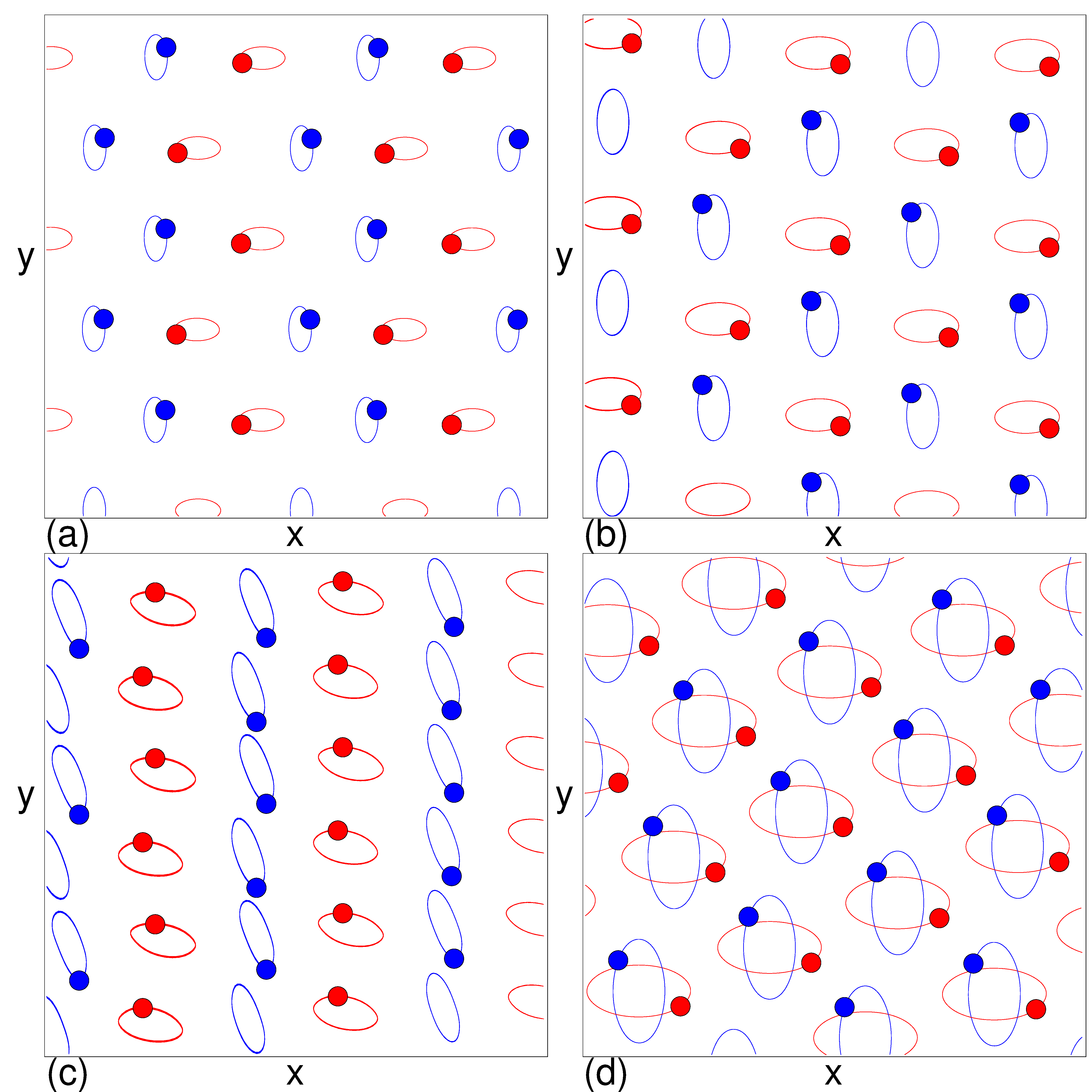}
\caption{Zoomed in view of particle positions (circles)
and trajectories (lines)
where species A is blue and species B is red for the out of phase
elliptically driven Coulomb particles
at $\rho=0.208$. As in Fig.~\ref{fig:1}(d),
the species A ellipses are aligned in the $y$ direction
and the species B ellipses are aligned in the $x$ direction.
(a) A distorted triangular crystal at
$\omega = 0.02$.
(b) A square spin ice lattice at
$\omega = 0.00145$.
(c) Elongated elliptical motion at
$\omega = 0.0125$.
(d) A paired crystal with a cross shape at 
$\omega = 0.008$.
}
\label{fig:18}
\end{figure}

We next demonstrate how to realize
a rich variety of other kinds of phase time crystals
as well as what we call phase time molecular crystals,
and show that distinct types of paired crystals can occur.
We specifically consider the case where the two species have the same
chirality but are driven out of phase by $180^\circ$
and have different elliptical driving
orbits, as shown in Fig.~\ref{fig:1}(d).
The chiral driving on species A is
$ {\bf F}^{A}(t)=A\sin(\omega t)\hat{\bf y}+B\cos(\omega t)\hat{\bf x}$,
and on species B it is
$ {\bf F}^{B}(t)= -B\sin(\omega t)\hat{\bf y} - A\cos(\omega t)\hat{\bf x}$.
For $A = 2.0$ and $B = 1.0$, species $A$ executes
an elliptical orbit with its long axis aligned in
the $y$ direction, while species B rotates
in an elliptical orbit aligned with the $x$ direction.
Under this out of phase elliptical driving,
we observe variations of the same phases found for out of phase
circular driving.
In Fig.~\ref{fig:18}(a) we show a subsection of the
$\omega = 0.02$ and $\rho = 0.208$ system 
where the orbits are small enough that a distorted triangular lattice (I)
state appears even though the individual particle orbits are clearly
elliptical.
At $\omega=0.0145$ in Fig.~\ref{fig:18}(b),
we find
what we call a spin ice-like lattice,
while in Fig.~\ref{fig:18}(c) a variation of this lattice appears at
$\omega = 0.0125$.
The terminology
``spin ice lattice'' is selected
in analogy to two-dimensional particle-based artificial spin ices
in which particles are placed in elongated traps that have
alternating alignment along the $x$ and $y$ directions
\cite{Nisoli13,Ambriz19}.
Figure~\ref{fig:18}(d) shows that at $\omega = 0.008$,
there is a paired crystal state in which
the elliptical orbits overlap
to create a cross-like geometry.
For out of phase circular orbits
with a drive amplitude of $A=1.0$ at
$\omega = 0.008$ and $\rho=0.208$,
the phase diagram in Fig.~\ref{fig:14} indicates that
a non-overlapping stripe lattice appears.
For the out of phase elliptical
orbits in Fig.~\ref{fig:18}(d), the larger ac amplitude along one direction
for each particle makes the orbits effectively larger, permitting
pairing to persist up to higher driving frequencies compared to
circular orbits.

\begin{figure}
\includegraphics[width=\columnwidth]{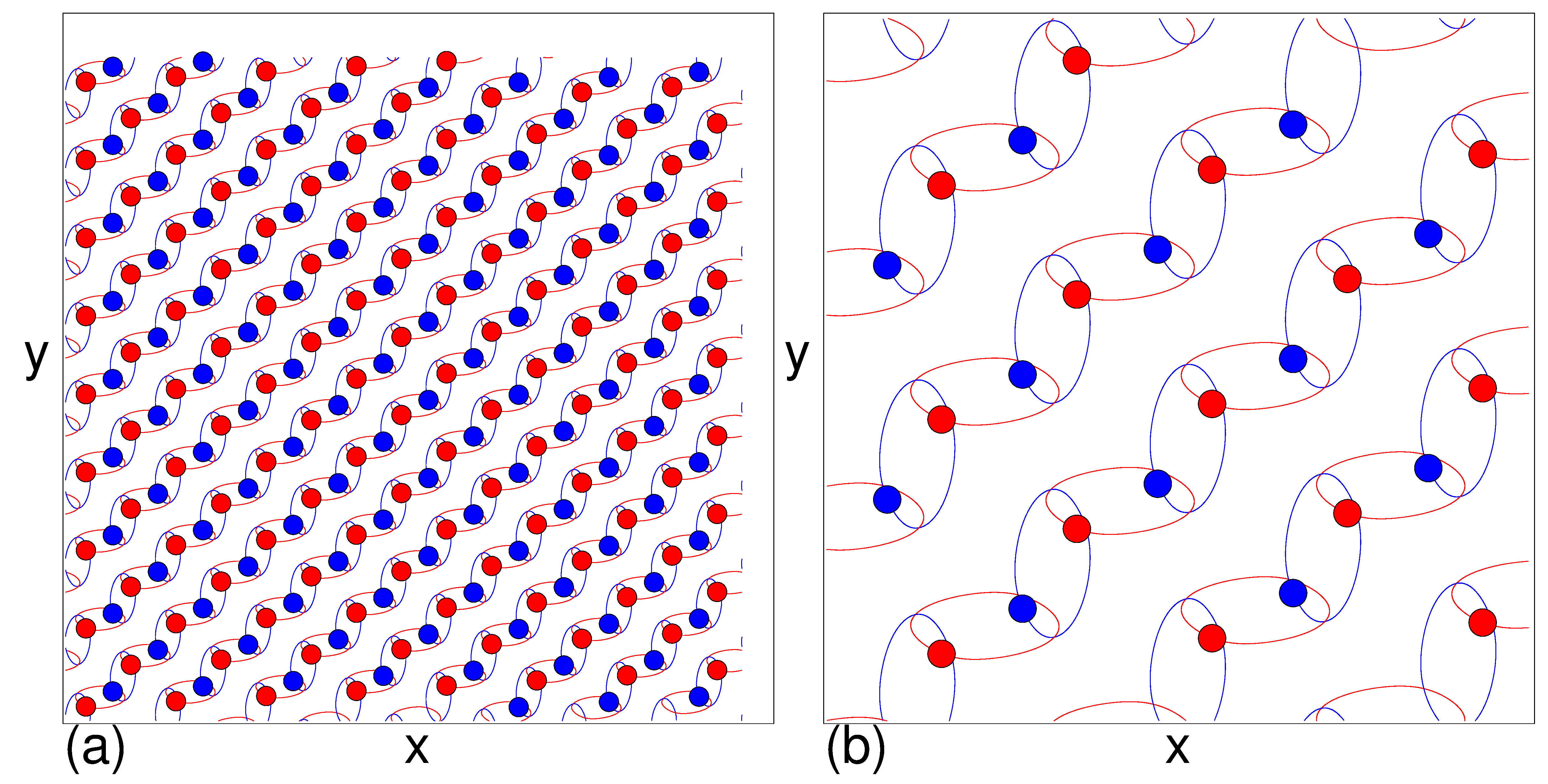}
\caption{(a) Particle positions (circles) and trajectories (lines)
where species A is blue and species B is red for counter-rotating
elliptically
driven Coulomb particles at $\rho = 0.208$ and $\omega = 0.008$,
showing a paired crystal with a cross shape.
(b) A zoom in of panel (a).}
\label{fig:19}
\end{figure}

If we consider particles with opposite chirality and elliptical drives,
the paired states are no longer present. Instead we find
chain crystal states of the type shown
in Fig.~\ref{fig:19}(a) at $\rho = 0.208$ and $\omega = 0.008$.
A zoomed in view of this state appears in
Fig.~\ref{fig:19}(b).
This state has similarities to the chain state that appears for
counter-rotating particles with circular rather than elliptical driving.

\begin{figure}
\includegraphics[width=\columnwidth]{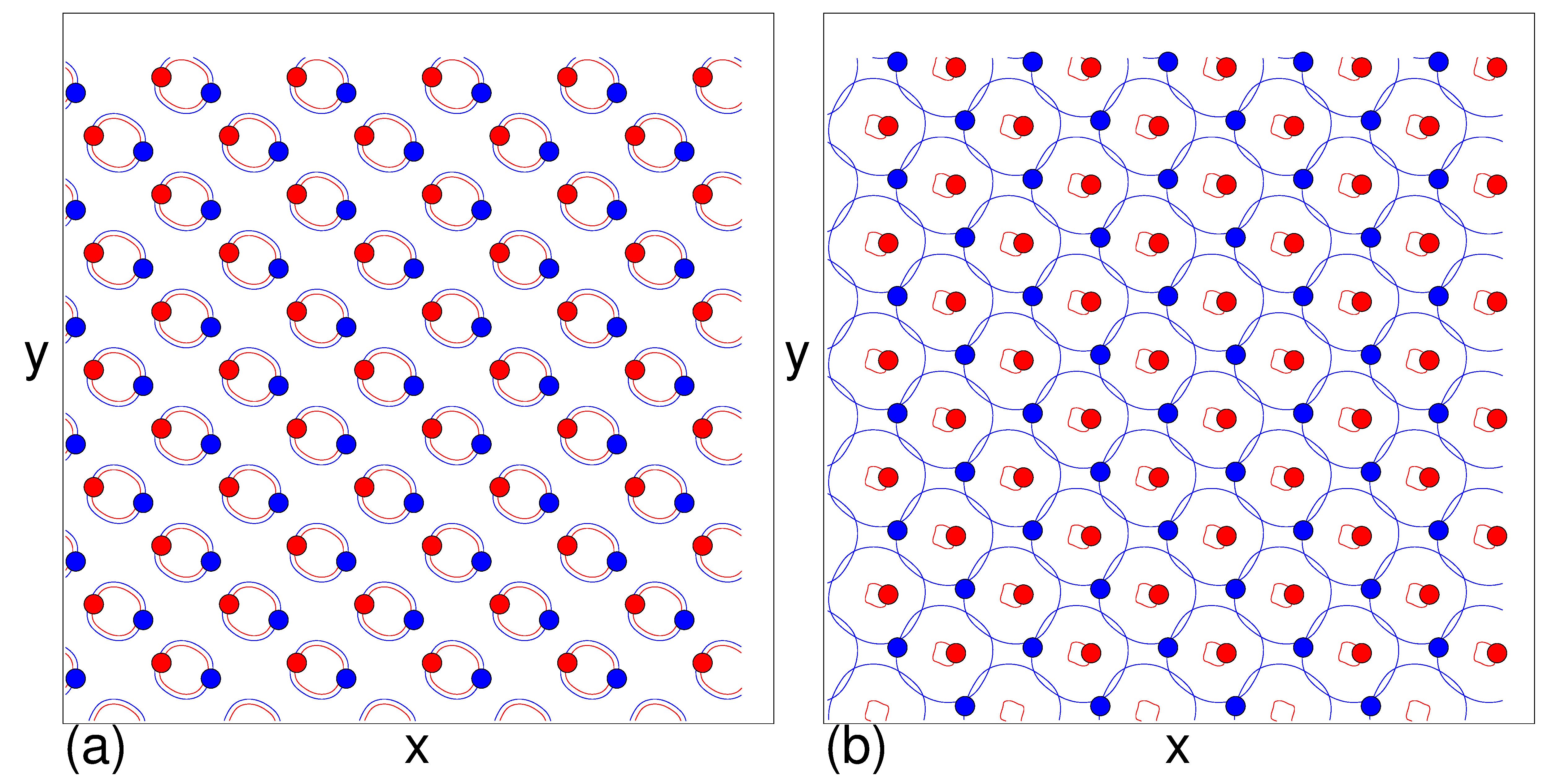}
\caption{Particle positions (circles) and trajectories (lines)
where species A is blue and species B is red for out of phase
Coulomb particles where the driving is circular but of unequal amplitude
in the two species.  
(a) $\rho = 0.208$ and $\omega = 0.006$,
where species $A$ has a drive amplitude of $A_A=1.2$ and species
$B$ has a drive amplitude of $A_B=1.0$.
(b) $\rho = 0.098$ and $\omega = 0.00375$,
where species $A$ has a drive amplitude of $A_A=2.0$
and species $B$ has a drive amplitude of $A_B=0.5$.
}
\label{fig:20}
\end{figure}

To further test the robustness of the pairing, we consider
a system with circular orbits of different sizes. The drive amplitude
is $A_A=1.2$ for species A and $A_B=1.0$ for species B, so that the two
species execute orbits of different radii.
It is still possible for 
a paired crystal state to form even for uneven orbit sizes, as shown
in Fig.~\ref{fig:20}(a) at $\rho = 0.208$ and $\omega = 0.006$,
where both particle species share a common orbit center point in each pair.
Figure~\ref{fig:20}(b)
shows the same system at $\rho = 0.098$ and $\omega = 0.004$ where
species A has a drive amplitude of $A_A=2.0$
and species B has an amplitude of $A_B=0.5$.
In this case, the inner orbit of the species B particles
has a square symmetry, which corresponds to the 
square symmetry of the large-scale lattice.

\begin{figure}
\includegraphics[width=\columnwidth]{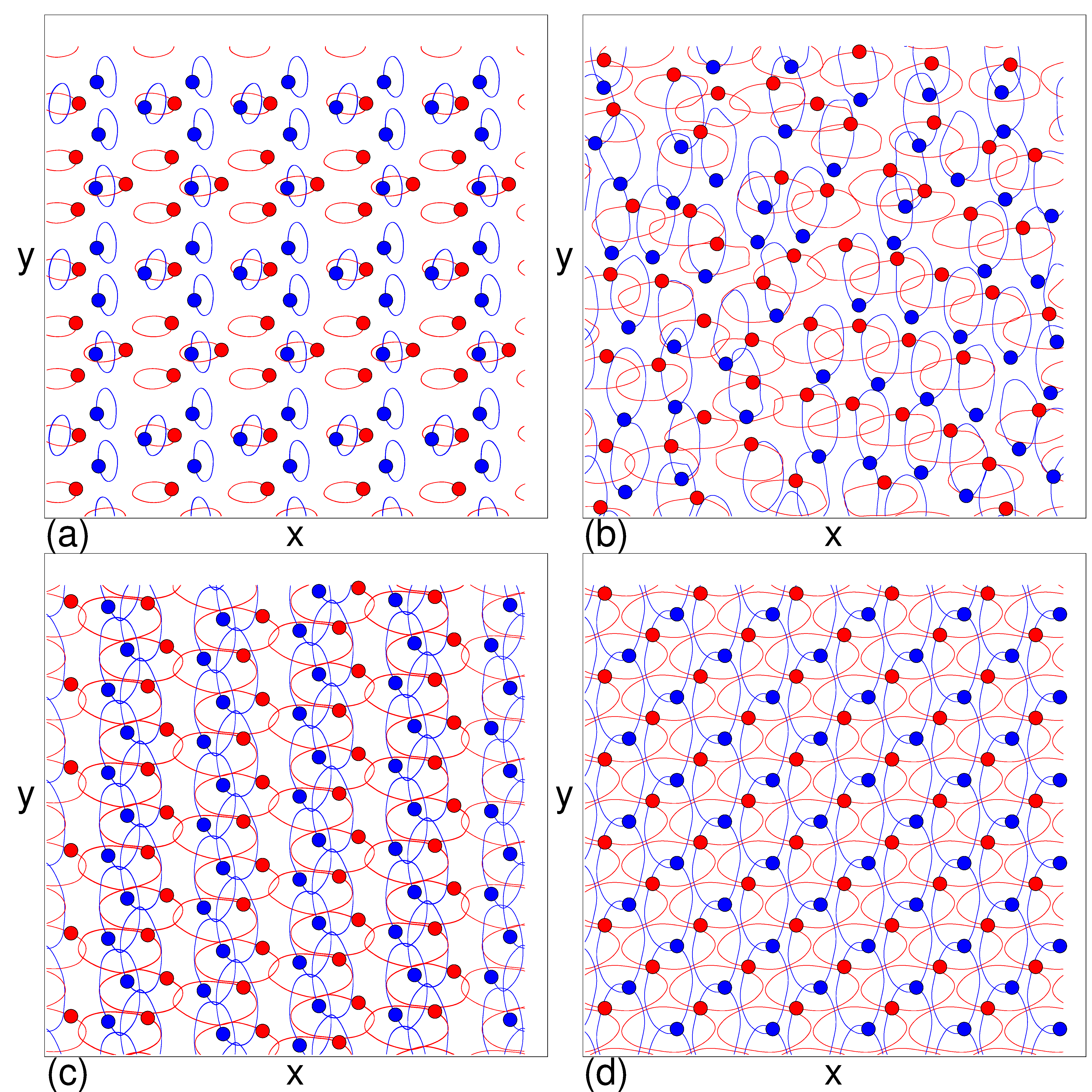}
\caption{Particle positions (circles) and trajectories (lines)
where species A is blue and species B is red for out of phase
elliptically driven Coulomb particles with $A=1.0$ and $B=2.0$ for
both particle species.
The species A orbits are aligned in the $y$ direction
and the species B orbits are aligned in the $x$ direction.
(a) A superlattice ordering of monomers and paired
states at $\rho = 0.098$ and $\omega = 0.00675$.
(b) A stripe glass state at $\rho = 0.098$ and $\omega = 0.00375$.
(c) A stripe crystal at $\rho = 0.098$ and $\omega = 0.00325$.
(d) An overlapping packed crystal
at $\rho = 0.098$ and $\omega = 0.003$.
}
\label{fig:21}
\end{figure}

For out of phase particles with the same chirality under
elliptical driving, we find additional phases at low particle densities.
In Fig.~\ref{fig:21}, where
$A = 1.0$ and $B = 2.0$ for both species, at
$\rho = 0.098$ and high $\omega$
we observe a distorted triangular crystal state
with non-overlapping orbits similar to that shown in
Fig.~\ref{fig:18}(b).
When we lower the driving frequency,
we observe superlattice ordering of paired and unpaired particles
as illustrated in Fig.~\ref{fig:21}(a) at $\omega = 0.00675$.
In this state,
every other row of paired particles has an alternating tilt,
and every paired particle is surrounded by six monomers.
For $0.004 < \omega < 0.00675$ we observe a paired cross crystal,
while for $\omega = 0.00375$ the system forms a stripe glass
as shown in Fig.~\ref{fig:21}(b).
At $\omega = 0.00325$ we find the stripe
crystal shown in Fig.~\ref{fig:21}(c),
while for even lower
$\omega=0.003$ the overlapping packed crystal state shown
in Fig.~\ref{fig:21}(d) appears.
If we further lower the frequency, we find
some additional regimes of overlapping packed crystals
and mixed fluids.

\begin{figure}
\includegraphics[width=\columnwidth]{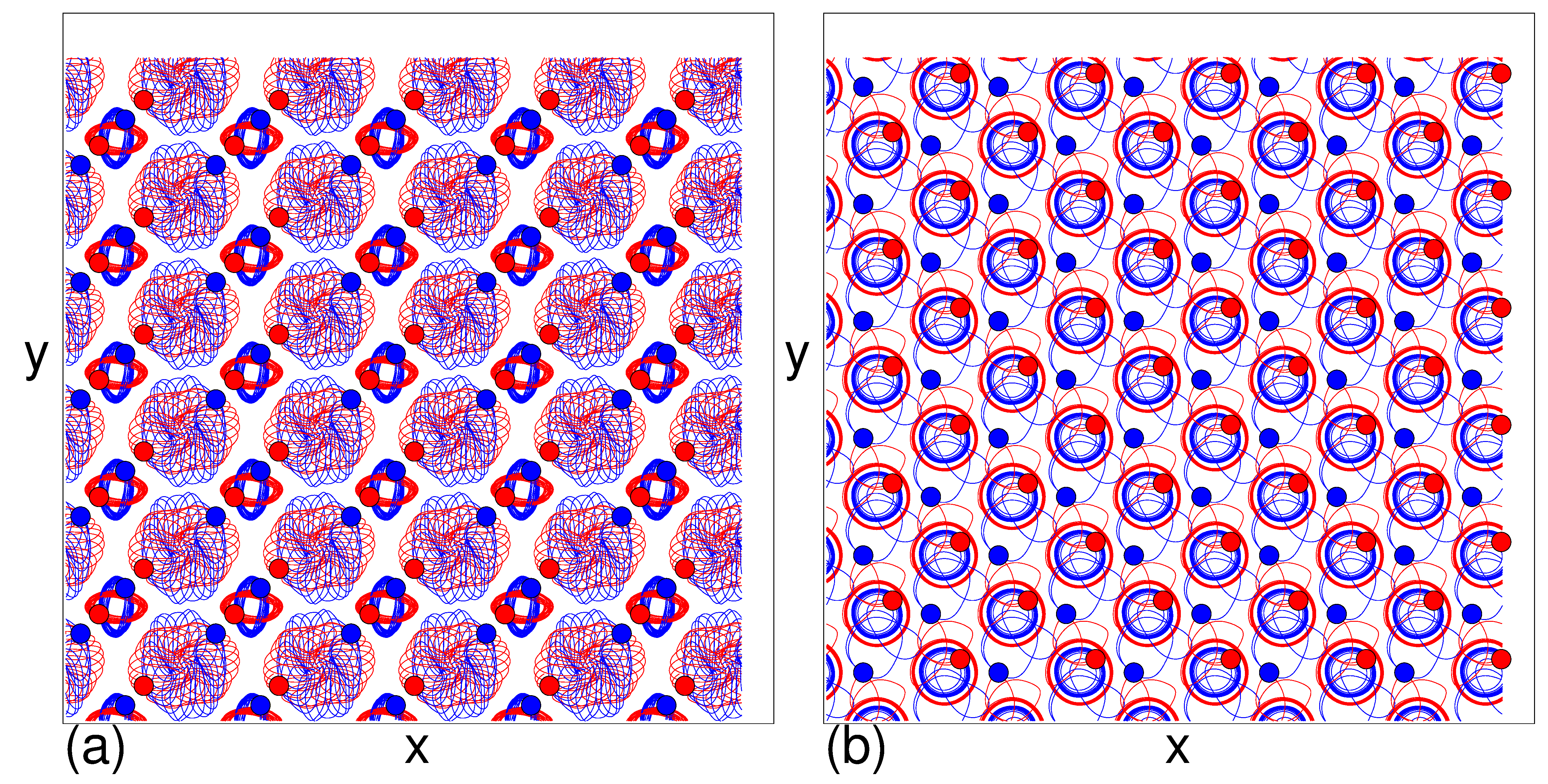}
\caption{Particle positions (circles) and trajectories (lines)
where species A is blue and species B is red for out of phase
Coulomb particles at $\rho=0.098$ showing transient behavior for
elliptical or circular driving.
(a) A system with elliptical driving
where $A = 2.0$ and $B = 1.0$, and where
$\omega = 0.007$ for both species,
showing the transient motion of a mixture of monomers
and a paired state.
(b) A system with circular driving at $A=1.0$
but different driving frequencies of
$\omega_A = 0.00348$ for species A and $\omega_B=0.0035$
for species B. The system forms a transient paired crystal before
jumping into a monomer state. The image shows the jumping process.}
\label{fig:22}
\end{figure}

For certain parameter windows, there can be precession of the
particle orbits.
The precession is purely transient when
both species are driven at the same frequency,
as has been considered throughout this work.
This is illustrated in Fig.~\ref{fig:22}(a)
for $\rho=0.098$ and $\omega=0.007$
in the same out of phase elliptically driven Coulomb particle system
from Fig.~\ref{fig:21}.
Here the monomers undergo large scale rotation around
the paired crosses until the system eventually
settles into the superlattice state
shown in Fig.~\ref{fig:21}(a).
If the driving frequencies of the two species
are not exactly matched,
we can also see transient states that have additional motion.
For example, if we consider out of phase circularly driven Coulomb particles,
as shown in Fig.~\ref{fig:14} there is a paired crystal state
at $\rho=0.098$ spanning the range $0.0027 < \omega < 0.005$.
We can drive each species at a frequency chosen from this window such
that the driving frequencies are similar but not identical.
In Fig.~\ref{fig:22}(b) we show the case where species A is driven
at $\omega_A=0.00348$ while species B is driven at $\omega_B=0.0035$.
Although the frequencies are relatively close together,
they are not close enough to
stabilize the paired crystal state. At short times the system enters the
paired crystal configuration,
but at some point the particles jump out of this state and form
a disordered monomer state. The trajectories during this transition process
are illustrated in Fig.~\ref{fig:22}(b).
We have considered other combinations of driving frequencies
and find that when $\omega_A/\omega_B$
is an integer or rational ratio,
it is possible to stabilize a paired state, while in other cases,
there can be a periodic transition between disordered and partially
ordered states over time.
These additional frequency effects will be explored in another work.

\section{Discussion}

The pairwise particle-particle interactions we consider in this work
favor formation of a triangular lattice in a static system.
The pattern formation we observe
arises through the competition of the pairwise interaction and
the activity of the driven particles.
One way to understand this 
is that the activity creates a time-averaged
flattening of the effective pairwise interactions.
For example, it is known that in systems where the interactions are
purely repulsive but where there are at least two length scales present,
such as through flattening of the interaction at some radius or through
the presence of a step in the potential,
numerous mesoscale patterned states can appear
such as bubbles or stripes \cite{Malescio03,Glaser07}. 
Such shoulder or core-corona type interaction potentials have
been proposed for describing
stripe and bubble states
of electrons in magnetic fields \cite{Fogler96}.
Bubble and stripe states
can also occur when the
particle interactions have
competing short-range attraction and
long-range repulsion 
terms \cite{Sciortino04,Reichhardt10,Hooshanginejad24}.

Several of the states we observe may represent
examples of transitions from a fluctuating state to a
non-fluctuating or absorbing state,
or to a periodically reversible state \cite{Reichhardt23}.
In a study of active circularly moving disks,
Lei {\it et al.} \cite{Lei19}
showed that the system starts off in a
fluctuating state but over time
either reaches a dynamically frozen state or
remains in a fluctuating state, so that
the frozen state is an example of an absorbing state.
For periodically sheared disks, the transition from
a fluctuating state to a cyclic reversible state
as a function of shear amplitude and/or particle density
is known as random organization \cite{Corte08}.
Similar transitions from fluctuating to reversible states
have also been demonstrated
for strongly interacting 
periodically sheared amorphous solids \cite{Regev13,Keim19},
indicating that absorbing phases can
appear in strongly interacting systems.
In our system, the glassy frozen states
typically have long transient times during which the system transitions
from a fluctuating state into a dynamically frozen state.
These glassy frozen states
would be the most similar to the dynamical frozen states
for the chiral system described by Lei {\it et al.} \cite{Lei19};
however, since our system has long-range repulsion, the
structures we observe are more stripe-like.

Our system can be viewed as
an extension of the concept of a time crystal
in that the ordered structures
we find are periodic in both space and
time.
In equilibrium systems,
pattern formation arises as a result of minimization of the
interaction energy.
It is possible that in our driven system, the particles
not only attempt to minimize their interaction energy, but at the same time
organize into a
pattern that maximizes the dissipation,
since the ordered structures
allow the particles to move
at a speed close
to the maximum velocity permitted by the driving.
There have been proposals
that driven self assembly processes can be understood as states
that maximize the dissipation \cite{England15}.
Futures directions could be to
study the dissipation of the different configurations
and determine whether the dissipation changes if the particles are forced
to move in specified orbits rather than the orbits into which they
naturally organize.

Other open questions include the nature of thermal melting transitions of
the dynamical states, whether the states
exhibit hysteresis across the boundaries of the phase diagrams
in Figs.~\ref{fig:8} and \ref{fig:14}, and what the nature of the
transitions among different pairs of phases is.
Some of the states we have described have similarities
to magnetic or particle-based artificial spin ice configurations,
so it could be interesting to explore the possible role of frustration,
and whether it is the strongly frustrated cases that form mixed fluids.
We considered two intermediate range and one long range particle-particle
interaction potential, but it would be possible to explore other forms
of interactions,
such as dipolar interactions or particles with competing attractive and
repulsive interactions.
In this work we only addressed 50:50 mixtures of the two species,
but it would be of interest to study different species ratios or
the role of dopants, such as by switching the species identity of a
small number of particles away from the 50:50 filling.
It could also be interesting to introduce
more than two particle species to see
whether it is still possible for the system to form frozen states.

There are a number of possible ways to realize
experimentally the dynamics we consider
using particles that have
an intermediate to long-range repulsion combined with some form of
circular motion.
These include spinning or circularly moving particles on the surface of a
fluid, which are known to undergo pattern formation \cite{Grzybowski01,Goto15}.
Other examples include charged particles such as colloids
with added chiral activity, charged particle systems on
periodic trap arrays where
the individual particle motion can be programmed,
robotic systems,
certain kinds of magnetic systems
in which there are mixtures of different topological textures,
and even coupled two-layer charge ordering systems
with one type of circular drive on one layer and another
drive on the other layer.

\section{Summary}

We have proposed a new type of chiral system we call phase time crystals,
which consists of two particle species with intermediate-to-long-range
repulsion that undergo circular motion of the same chirality but
out of phase by $180^\circ$ between the species.
We study a two-dimensional system of Bessel function, Yukawa, and
Coulomb interacting particles that have
activity in the form of circular driving.
As a function of the driving frequency, which changes the size of the
orbit, and the particle density,
we find a remarkable variety of stable dynamical crystalline states.
For high frequencies when the orbits are much smaller than the
equilibrium particle spacing, there is
a distorted triangular lattice.
As the frequency decreases, the orbit size increases
and the system forms a stripe-like crystal that becomes increasingly
disordered as the orbits distort.
A phase separation occurs when the frequency decreases further,
allowing the particles to follow
circular orbits that do not overlap.
Once the frequency becomes low enough that the orbit of an individual particle
is slightly larger
than the average spacing between the static particles,
a pairing of out of phase particles occurs in which each orbit contains
one particle of each phase moving on opposite sides of the orbit, and
the pairs themselves
order into a crystalline state.
As the orbit size continues to increase with decreasing frequency, the
paired crystal destabilizes into a phase-separated fluid state,
and near certain ratios of the equilibrium lattice constant
to the orbit size,
we observe overlapping packed crystals and overlapping stripe crystals.
At the lowest frequencies, the system forms a mixed fluid.
We show that these states, including the paired crystal,
are robust for a wide range of orbit sizes and densities
as well as for all of the different particle interactions we consider.

For low-density systems where the effective
particle-particle interactions are weaker,
we find disordered states in which the individually orbiting particles
have no long-range organization
but the system is dynamically frozen in that there is no long-time diffusion.
We show that the states we observe are robust against thermal fluctuations.
In particular, the paired crystal persists
up to the same temperature at which the non-driven crystal melts,
and the pairs themselves are able to persist at higher temperatures in
the form of a paired fluid.
If we initialize
the system in a random state,
the transient times increase but the same phases still occur.
If we change the driving so that the particles are counter-rotating instead
of out of phase,
the paired state does not occur and is replaced by
chain lattice states where the particle species alternate
along the chains. This provides a similar reduction in the pairwise
interaction energy throughout the ac cycle as the out of phase particles
achieved in the paired crystal state.
If we use elliptical rather than circular driving for the out of phase
particles, where the orientation of the ellipses is different for each
species, we find
a series of more complex states
that we call molecular time crystals, which include states
with spin ice ordering and states with superlattice ordering.
The paired crystal states can still occur for elliptical driving,
but the orbits only partially overlap to form a cross-like configuration.

We discuss how the states we observe can be considered
as an extension of a time crystal, since
they are periodic in both space and time.
The system can be viewed as simultaneously
minimizing the repulsive pairwise potential energy
while also maximizing the dissipation or particle velocity.
We also discuss how the organization of the system into
dynamical frozen disordered states
could be an example of an absorbing phase transition
similar to the transitions to reversible states
found in cyclically driven systems.
Our results could be realized in systems with charge ordering,
charged colloids, charged or magnetic particles
at interfaces, bilayer systems,
magnetic textures where there is also some
form of circular driving, robotic systems,
and mixtures of magnetic and charged particles.

\begin{acknowledgements}
We gratefully acknowledge the support of the U.S. Department of
Energy through the LANL/LDRD program for this work.
This work was supported by the US Department of Energy through
the Los Alamos National Laboratory.  Los Alamos National Laboratory is
operated by Triad National Security, LLC, for the National Nuclear Security
Administration of the U. S. Department of Energy (Contract No. 892333218NCA000001).
\end{acknowledgements}

\bibliography{mybib}

\end{document}